\documentclass[pre,11pt,showpacs]{revtex4}
\usepackage{amsmath}
\usepackage{graphicx}
\usepackage{subfigure}
\usepackage{amssymb}
\usepackage{subfig}

\input{tcilatex}
\begin{document}

\title{A solvable model for solitons pinned to a $\mathcal{PT}$-symmetric
dipole}
\author{Thawatchai Mayteevarunyoo$^{1}$, Boris A. Malomed$^{2}$ and Athikom
Reoksabutr$^{1}$}
\affiliation{$^{1}$Department of Telecommunication Engineering, Mahanakorn University of
Technology, Bangkok 10530, Thailand\\
$^{2}$Department of Physical Electronics, School of Electrical Engineering,
Faculty of Engineering, Tel Aviv University, Tel Aviv 69978, Israel}

\begin{abstract}
We introduce the simplest one-dimensional nonlinear model with the
parity-time ($\mathcal{PT}$) symmetry, which makes it possible to find exact
analytical solutions for localized modes (\textquotedblleft solitons"). The $%
\mathcal{PT}$-symmetric element is represented by a point-like
(delta-functional) \textit{gain-loss dipole} $\sim \delta ^{\prime }(x)$,
combined with the usual attractive potential $\sim \delta (x)$. The
nonlinearity is represented by self-focusing (SF) or self-defocusing (SDF)
Kerr terms, both spatially uniform and localized ones. The system can be
implemented in planar optical waveguides. For the sake of comparison, also
introduced is a model with separated $\delta $-functional gain and loss,
embedded into the linear medium and combined with the $\delta $-localized
Kerr nonlinearity and attractive potential. Full analytical solutions for
pinned modes are found in both models. The exact solutions are compared with
numerical counterparts, which are obtained in the gain-loss-dipole model
with the $\delta ^{\prime }$- and $\delta $- functions replaced by their
Lorentzian regularization. With the increase of the dipole's strength, $%
\gamma $, the single-peak shape of the numerically found mode, supported by
the uniform SF nonlinearity, transforms into a double-peak one. This
transition coincides with the onset of the escape instability of the pinned
soliton. In the case of the SDF uniform nonlinearity, the pinned modes are
stable, keeping the single-peak shape.
\end{abstract}

\maketitle

\section{Introduction}

Recently, a great deal of interest has been drawn to physical systems
featuring the $\mathcal{PT}$ (parity-time) symmetry \cite%
{Bender_review,special-issues,review}, i.e., dissipative quantum or wave
systems with the antisymmetry between spatially separated gain and loss. In
particular, making use of the similarity of the quantum-mechanical Schr\"{o}%
dinger equation to the parabolic propagation equation in optics, it was
proposed theoretically~\cite{Muga} and demonstrated experimentally \cite%
{experiment} that the $\mathcal{PT}$ symmetry can be realized in the
classical context of the optical wave propagation, using waveguides with the 
$\mathcal{PT}$-balanced gain and loss. Very recently, an experimental
realization of the $\mathcal{PT}$\ symmetry was also reported in a system of
coupled electronic oscillators \cite{Kottos}. These findings have stimulated
numerous studies of the linear wave propagation in $\mathcal{PT}$-symmetric
systems \cite{special-issues}, especially in those including periodic
potentials~\cite{PT_periodic} (see also review~\cite{review}).

The optical realizations of the $\mathcal{PT}$ symmetry suggest additional
interest to nonlinearity in these systems \cite{Musslimani2008}, in which
stable solitons can be supported by the combination of the Kerr (cubic)
nonlinearity and spatially periodic complex potential, whose odd
(antisymmetric) imaginary part accounts for the balanced gain and loss, as
mentioned above. Stability of such solitons was rigorously analyzed in Ref.~%
\cite{Yang}. Dark solitons were also investigated in in the framework of
models combining the $\mathcal{PT}$ symmetry and self-defocusing Kerr
nonlinearity \cite{dark}. In addition to that, bright solitons were
predicted too in $\mathcal{PT}$-symmetric systems with the quadratic
(second-harmonic-generating) nonlinearity \cite{chi2}.

Solitons can also be found in linearly-coupled dual-core systems, with the
balanced gain and loss acting in the two cores, and the intrinsic Kerr
nonlinearity present in each one \cite{dual,dual2}. Further, discrete
solitons were predicted in various models based on chains of linear \cite%
{discrete} and circular \cite{circular} coupled $\mathcal{PT}$-symmetric
elements and, more generally, in networks of coupled $\mathcal{PT}$%
-symmetric oligomers (dimers, quadrimers, etc.)~\cite{KPZ}. Parallel to
incorporating the usual Kerr nonlinearity into the conservative part of the
system, its gain-loss-antisymmetric part can be made nonlinear too, by
introducing mutually balanced cubic gain and loss terms ~\cite{AKKZ}.
Effects of combined linear and nonlinear $\mathcal{PT}$~terms on the
existence and stability of optical solitons were studied too \cite{combined}.

Unlike the usual nonlinear dissipative systems, where solitons exist as
isolated solutions (\textit{attractors}) \cite{PhysicaD,Kutz}, in $\mathcal{%
PT}$-symmetric settings solitons form continuous families, similar to the
generic situation in conservative media. However, the increase of the
gain-loss coefficient ($\gamma $) in the $\mathcal{PT}$-symmetric nonlinear
system leads to shrinkage of existence and stability areas for $\mathcal{PT}$%
-symmetric solitons, until they vanishes when this coefficient attains a
critical value, $\gamma _{\mathrm{cr}}$.

In all the previously studied models of $\mathcal{PT}$-symmetric nonlinear
systems, except for the simplest dual-core model considered in Ref. \cite%
{dual}, solitons could only be constructed and investigated in a numerical
form (in Ref. \cite{dual}, the solutions for $\mathcal{PT}$-symmetric
solitons were tantamount to those for symmetric solitons in the usual
coupler model, the difference being in their stability). The objective of
the present work is to propose a \textit{solvable} one-dimensional nonlinear
model with the $\mathcal{PT}$-balanced gain and loss concentrated at a
single point, in the form of a $\mathcal{PT}$ \textit{dipole}. A possibility
to construct such a tractable model is suggested by recently studied models
of dissipative systems (not subject to the condition of the $\mathcal{PT}$
symmetry), in which localized gain, competing with spatially uniform loss,
was applied at a single \cite{singleHS} or two \cite{twoHS} ``hot spots",
making it possible to find exact solutions for dissipative solitons pinned
to those spots. Similar models, but with hot spots of a finite width, were
investigated by means of numerical methods in Refs. \cite{finite-width}.

A $\mathcal{PT}$-symmetric model with the uniformly distributed nonlinearity
and localized mutually balanced gain and loss, applied at two points in the
form of $\delta $-functions, along with the attractive potential,
represented by a pair of $\delta $-functions placed at the same points, was
elaborated in Ref. \cite{Wunner}. The model dealt with in the present work
may be considered as a limit form of the one introduced in Ref. \cite{Wunner}%
, for a vanishingly small separation between the two $\delta $-functions,
when the balanced pair of $\delta $-function-shaped gain and loss go over
into a single term in the underlying propagation equation, represented by
the $\delta ^{\prime }$-function (the ``$\mathcal{PT}$ dipole").

The model with the $\mathcal{PT}$ dipole embedded into the Kerr-nonlinear
medium admits a full family of analytical solutions for localized modes
(``solitons") pinned to the point-like dipole (``defect"), for both signs of
the nonlinearity of the host medium, self-focusing (SF) and self-defocusing
(SDF). Unlike what occurs in other $\mathcal{PT}$-symmetric systems, the
analytical solutions exist at all values of the gain-loss parameter, $\gamma 
$, without featuring the above-mentioned threshold (critical value), $\gamma
_{\mathrm{cr}}$. On the other hand, our numerical solution for the model
with the $\delta $- and $\delta ^{\prime }$- functions replaced by their
finite-width regularizations [see Eq. (\ref{Lorentz}) below] demonstrate
that, while at $\gamma $ small enough the numerical solutions are close to
their analytical counterparts found for the ideal $\delta $-functions, their
stability and existence are always limited by finite $\gamma _{\mathrm{cr}}$.

The appearance of the threshold with the introduction of the regularization,
i.e., finite separation between the gain and loss, can also be explained in
an analytical form. To this end, we introduce an additional model, based on
the linear host medium with an embedded pair of \emph{separated} $\delta $%
-functions, which carry, in addition to the gain and loss with equal
coefficients $\Gamma $, the local cubic SF nonlinearity, along with the
linear $\delta $-functional potentials. This setting also admits a full
analytical solution (cf. Refs. \cite{Dong} and \cite{Yasha}, where exact
solutions were found for one- and two-component symmetric, antisymmetric,
and asymmetric solitons pinned to a pair of points with the localized SF
nonlinearity, embedded into the linear medium). Analytical solutions
obtained in the system with the separated $\delta $-functions explicitly
feature a threshold value, $\Gamma _{\mathrm{cr}}$, which bounds their
existence region.

The solvable models and analytical solutions are introduced in Section II.
Their numerically found counterparts, corresponding to the regularized $%
\delta ^{\prime }$ and $\delta $- functions, are presented in Section III.
The numerical analysis pursues two objectives: to estimate the robustness
(structural stability) of the analytical solutions for the localized modes,
obtained with the ideal $\delta ^{\prime }$- and $\delta $- functions, and
to test the dynamical stability of the modes by means of systematic
simulations of the perturbed evolution, which is a crucially important issue
in the context of $\mathcal{PT}$-symmetric systems. The paper is concluded
by Section IV.

\section{The model and analytical solutions}

\subsection{The basic model}

The underlying equation for the light propagation in a nonlinear planar
waveguide, with the $\mathcal{PT}$-symmetric complex potential concentrated
near $x=0$ (with even real and odd imaginary parts), and the uniform cubic
nonlinearity with coefficient $\sigma $, is%
\begin{equation}
iu_{z}=-\frac{1}{2}u_{xx}-\left( \varepsilon _{0}+\varepsilon
_{2}|u|^{2}\right) u\delta \left( x\right) +i\gamma u\delta ^{\prime }\left(
x\right) -\sigma |u|^{2}u.  \label{eq}
\end{equation}%
Here $z$ is the propagation distance and $x$ the transverse coordinate, with
term $u_{xx}$ accounting for the transverse diffraction in the paraxial
approximation, while $\varepsilon _{0}>0$ and $\gamma >0$ are strengths of
the real and imaginary parts of the complex potential, and $\varepsilon _{2}$
represents a possible nonlinear part of the trapping potential \cite{Dong}.
By means of obvious rescaling, one can set $\left\vert \sigma \right\vert
\equiv 1$, with $\sigma =+1$ and $-1$ corresponding, respectively, to the SF
an SDF spatially uniform nonlinearity. In addition, $\sigma =0$ is possible
too, corresponding to the model with the Kerr nonlinearity fully
concentrated at the same spot where the $\mathcal{PT}$ dipole is set.
Rescaling also allows us to fix $\varepsilon _{0}\equiv 1$, unless $%
\varepsilon _{0}=0$, in which case it is possible to fix $\gamma \equiv 1$.
The sign of $\varepsilon _{2}$ in Eq. (\ref{eq}) may be either the same as $%
\sigma =\pm 1$ or\textit{\ }opposite to it, the latter case corresponding to
the competition between the uniform and localized nonlinearities.

Stationary $\mathcal{PT}$-symmetric localized solutions to Eq. (\ref{eq})
are looked for as%
\begin{gather}
u\left( x,z\right) =e^{ikz}U(x),  \label{UU} \\
U^{\ast }(x)=U(-x).  \label{UUU}
\end{gather}%
with real propagation constant $k>0$, where complex function $U(x)$ obeys
the following equation:%
\begin{equation}
kU-\frac{1}{2}U^{\prime \prime }-\sigma |U|^{2}U-\left( \varepsilon
_{0}+\varepsilon _{2}|U|^{2}\right) U\delta \left( x\right) +i\gamma U\delta
^{\prime }\left( x\right) =0.  \label{U}
\end{equation}%
At $x\neq 0$, $\mathcal{PT}$-symmetric solutions of Eq. (\ref{U}) for
localized modes are constructed in terms of the commonly known analytical
expressions for regular and singular solitons of the nonlinear Schr\"{o}%
dinger (NLS) equation with the SF or SDF nonlinearity ($\sigma =+1$ and $%
\sigma =-1$, respectively): 
\begin{eqnarray}
U(x) &=&\sqrt{2k}\frac{\cos \theta +i~\mathrm{sgn}(x)\sin \theta }{\cosh
\left( \sqrt{2k}\left( |x|+\xi \right) \right) },\mathrm{~for}~~\sigma =+1,
\label{SF} \\
U(x) &=&\sqrt{2k}\frac{\cos \theta +i~\mathrm{sgn}(x)\sin \theta }{\sinh
\left( \sqrt{2k}\left( |x|+\xi \right) \right) },~\mathrm{for}~~\sigma =-1,
\label{SDF}
\end{eqnarray}%
where $\theta $ and $\xi $ are free real parameters. The form of this
solution implies that $\mathrm{Im}\left( U(x=0)\right) =0,$ while jumps ($%
\Delta $) of the imaginary part and first derivative of the real part at $%
x=0 $ are determined by the integration of the $\delta $- and $\delta
^{\prime }$- functions in an infinitesimal vicinity of $x=0$:%
\begin{eqnarray}
\Delta \left\{ \mathrm{Im}\left( U\right) \right\} |_{x=0} &=&2\gamma _{0}%
\mathrm{Re}\left( U\right) |_{x=0},  \label{Im} \\
\Delta \left\{ \left( \frac{d}{dx}\mathrm{Re}\left( U\right) \right)
\right\} |_{x=0} &=&-2\left[ \varepsilon _{0}+\varepsilon _{2}\left( \mathrm{%
Re}\left( U\right) \right) ^{2}\right] \mathrm{Re}\left( U\right) |_{x=0}.
\label{Re}
\end{eqnarray}

\subsection{The analytical solution for the model with the spatially uniform
nonlinearity ($\protect\varepsilon _{2}=0$)}

Substituting bulk solutions (\ref{SF}) and (\ref{SDF}) into boundary
conditions (\ref{Im}) and (\ref{Re}) with $\varepsilon _{2}=0$, it is
straightforward to obtain the following results, which determine the free
constants in the solutions, $\theta $ and $\xi $, as functions of $k$:%
\begin{equation}
\theta =\arctan \left( \gamma \right) ,  \label{theta}
\end{equation}%
which does not depend on $k$ and is the same for $\sigma =\pm 1$, and%
\begin{equation}
\xi =\frac{1}{2\sqrt{2k}}\ln \left( \sigma \frac{\sqrt{2k}+\varepsilon _{0}}{%
\sqrt{2k}-\varepsilon _{0}}\right) .  \label{xi}
\end{equation}%
The total power of the localized mode is 
\begin{equation}
P_{\sigma }\equiv \int_{-\infty }^{+\infty }|U(x)|^{2}dx=2\sigma \left( 
\sqrt{2k}-\varepsilon _{0}\right) .  \label{Power}
\end{equation}

As seen from Eq. (\ref{xi}), the solutions exist at 
\begin{equation}
\left\{ 
\begin{array}{cc}
\sqrt{2k}>\varepsilon _{0} & \mathrm{for}~~\sigma =+1, \\ 
\sqrt{2k}<\varepsilon _{0} & \mathrm{for~~}\sigma =-1.%
\end{array}%
\right. ~~  \label{><}
\end{equation}%
As concerns stability of the solutions, it is relevant to mention that
expression (\ref{Power}) with $\sigma =+1$ and $-1$ satisfy, respectively,
the Vakhitov-Kolokolov (VK) \cite{VK} and ``anti-VK" \cite{anti} criteria,
i.e., 
\begin{equation}
dP_{+1}/dk>0,~dP_{-1}/dk<0,  \label{Vakh}
\end{equation}%
which are necessary conditions for the stability of localized modes
supported, severally, by the SF and SDF nonlinearities, hence in both cases
the present solutions have a chance to be stable.

\subsection{Analytical solutions for the\ model with the inhomogeneous
nonlinearity ($\protect\varepsilon _{2}\neq 0$)}

In the presence of the localized nonlinearity, Eq. (\ref{theta}) remains the
same as before, while expression (\ref{xi}), following from Eq. (\ref{Re}),
is replaced by a rather cumbersome expression:%
\begin{gather}
\left[ \tanh \left( \sqrt{2k}\xi \right) \right] ^{\sigma }=-\frac{1+\gamma
^{2}}{2\sqrt{2k}\varepsilon _{2}}  \notag \\
\pm \sqrt{\frac{\left( 1+\gamma ^{2}\right) ^{2}}{8k\varepsilon _{2}^{2}}%
+1+\sigma \frac{\varepsilon _{0}\left( 1+\gamma ^{2}\right) }{\sqrt{2k}%
\varepsilon _{2}}}.  \label{tanh+}
\end{gather}%
These solutions may be free of singularities and stable only if they yield $%
\xi >0$. In the case of $\varepsilon _{2}>0$ (the attractive nonlinear
potential placed at $x=0$), the condition of $\xi >0$ for Eq. (\ref{tanh+})
with $\sigma =+1$ holds only with sign $+$ in front of the radical, while
Eq. (\ref{tanh+}) with $\sigma =-1$ may give rise to two different
solutions, corresponding to both signs $+$ and $-$. In the case of $%
\varepsilon _{2}<0$, the situation is opposite: Eq. (\ref{tanh+}) with $%
\sigma =-1$ makes sense only with sign $+$ in front of the radical, while $%
\sigma =+1$ may generate meaningful solutions for both signs $+$ and $-$.
Thus, two different solutions may exist in the case of the competition
between the uniform and localized nonlinearities.

It is also relevant to consider the special case of $\varepsilon _{0}=0$, $%
\varepsilon _{2}>0$, when the attractive potential of the defect at $x=0$ is
purely nonlinear. In this situation, solution (\ref{tanh+}) essentially
simplifies, taking the same form for $\sigma =\pm 1$:%
\begin{equation}
\xi =\frac{1}{2\sqrt{2k}}\ln \left[ \frac{2\varepsilon _{2}\sqrt{2k}}{%
1+\gamma ^{2}}+\sqrt{1+\frac{8\varepsilon _{2}^{2}k}{\left( 1+\gamma
^{2}\right) ^{2}}}\right] .  \label{varepsilon_0=0}
\end{equation}%
The corresponding expressions for the total power can be found too, cf. Eq. (%
\ref{Power}):%
\begin{equation}
P_{\sigma }(k)=2\left[ \frac{1+\gamma ^{2}}{2\varepsilon _{2}}+\sigma \sqrt{%
2k}-\sigma \sqrt{2k+\frac{\left( 1+\gamma ^{2}\right) ^{2}}{4\varepsilon
_{2}^{2}}}\right] .  \label{P-}
\end{equation}%
Note that this expression depends on $\gamma $, unlike its counterpart (\ref%
{Power}).

Solution (\ref{varepsilon_0=0}) exists for all values of $k>0$, unlike the
one given by Eq. (\ref{xi}), whose existence region is limited by condition (%
\ref{><}). Further, expression (\ref{P-}) satisfies the VK and anti-VK
criteria (\ref{Vakh}), severally for $\sigma =+1$ and $\sigma =-1$, hence in
both cases solution (\ref{varepsilon_0=0}) may be stable.

\subsection{The analytical solution for the linear host medium ($\protect%
\sigma =0$)}

For the nonlinear $\mathcal{PT}$ dipole embedded into the linear medium, it
is possible to fix $\varepsilon _{2}=\pm 1$, for the SF and SDF localized
nonlinearity, respectively. The solution of Eq. (\ref{U}) for the trapped
mode is simple in this case:%
\begin{equation}
U(x)=\sqrt{\frac{\sqrt{2k}-\varepsilon _{0}}{\varepsilon _{2}}}\left[
1+i\gamma \mathrm{sgn}(x)\right] e^{-\sqrt{2k}|x|},  \label{sigma=0}
\end{equation}%
with the corresponding total power%
\begin{equation}
P_{0}(k)=\frac{1+\gamma ^{2}}{\varepsilon _{2}}\left( 1-\frac{\varepsilon
_{0}}{\sqrt{2k}}\right) .  \label{P0}
\end{equation}%
Like the above solution given by Eq. (\ref{xi}), and unlike the one
amounting to Eq. (\ref{varepsilon_0=0}), the existence of this solution is
limited by conditions $\sqrt{2k}>\varepsilon _{0}$ and $\sqrt{2k}%
<\varepsilon _{0}$, respectively, in the case of the SF and SDF localized
nonlinearity, cf. Eq. (\ref{><}). Further, as well as the solution families
considered above, relation (\ref{P0}) satisfies the VK and anti-VK criteria
[see Eq. (\ref{Vakh})] for the SF and SDF signs of the localized
nonlinearity, i.e., $\varepsilon _{2}=+1$ and $\varepsilon _{2}=-1$. With $%
\varepsilon _{0}=0$, Eq. (\ref{P0}) yields the degenerate dependence, $%
dP_{0}/dk=0$, which formally implies VK-neutral stability, but in reality
the solitons are unstable in this case \cite{Dror}.

\subsection{The model with the separated gain and loss embedded into the
linear medium}

As explained above, it is relevant to supplement the $\mathcal{PT}$-dipole
model by a solvable one which features a finite separation, $2l$, between
the mutually balanced gain and loss $\delta $-like elements with equal
strengths $\Gamma $. Such a system may be built following the lines of Ref. 
\cite{Wunner}, but replacing the uniform Kerr nonlinearity by its
counterpart localized at the same points where the gain and loss are set
(otherwise, the system is not analytically solvable):%
\begin{gather}
iu_{z}=-\frac{1}{2}u_{xx}-\left( \varepsilon _{0}+\varepsilon
_{2}|u|^{2}\right) \left[ \delta \left( x-l\right) +\delta \left( x+l\right) %
\right] u  \notag \\
+i\Gamma \left[ \delta \left( x-l\right) -\delta \left( x+l\right) \right] u.
\label{Wunn}
\end{gather}%
Using obvious rescaling, we can fix here $\ \left\vert \varepsilon
_{2}\right\vert =1$ for the SF ($\varepsilon _{2}=+1$) and SDF ($\varepsilon
_{2}=-1$) signs of the nonlinearity.

Stationary $\mathcal{PT}$-symmetric localized solutions to Eq. (\ref{Wunn})
are looked in the same form (\ref{UU}) and (\ref{UUU}) as above, with $U(x)$
obeying equation%
\begin{gather}
kU-\frac{1}{2}U^{\prime \prime }-\left( \varepsilon _{0}+\varepsilon
_{2}|U|^{2}\right) \left[ \delta \left( x-l\right) +\delta \left( x+l\right) %
\right] U  \notag \\
+i\Gamma \left[ \delta \left( x-l\right) -\delta \left( x+l\right) \right]
U=0.  \label{Gamma}
\end{gather}%
At $|x|~>l$ and $|x|~<l$, respectively, $\mathcal{PT}$-symmetric solutions
of Eq. (\ref{U}) for $U(x)$ are constructed as follows, cf. Eq. (\ref%
{sigma=0}):%
\begin{gather}
U_{\mathrm{out}}(x)=\left[ A+i~\mathrm{sgn}(x)B\right] \exp \left( -\sqrt{2k}%
\left( |x|-l\right) \right) ,  \notag \\
U_{\mathrm{in}}(x)=\left[ C\cosh \left( \sqrt{2k}x\right) +iD\sinh \left( 
\sqrt{2k}x\right) \right] ,  \label{in-out}
\end{gather}%
where $A,B$ and $C,D$ are real amplitudes. The condition of the continuity
of solution (\ref{in-out}) at $x=\pm l$ yields a relation eliminating $C$
and $D$ in favor of $A$ and $B$:%
\begin{equation}
C\cosh \left( \sqrt{2k}l\right) +iD\sinh \left( \sqrt{2k}l\right) =A+iB.
\label{CDAB}
\end{equation}

The condition for the jump of the first derivatives induced by the
delta-functions at $x=\pm l$, cf. Eq. (\ref{Re}), can be written as a single
cubic complex equation for $A$ and $B$:%
\begin{gather}
\sqrt{2k}\left( \frac{A}{1+e^{-2\sqrt{2k}l}}+\frac{iB}{1-e^{-2\sqrt{2k}l}}%
\right)  \notag \\
=\left[ \varepsilon _{0}+\varepsilon _{2}\left( A^{2}+B^{2}\right) -i\Gamma %
\right] \left( A+iB\right) .  \label{AB}
\end{gather}

Equation (\ref{AB}) may be considered as a system of two homogeneous
equations for $A$ and $B$, a nontrivial solution to which exists when the
system's determinant vanishes. After some algebra, an explicit solution of
Eq. (\ref{AB}) can be obtained:%
\begin{equation}
A=\sqrt{\frac{Q-\varepsilon _{0}}{\varepsilon _{2}}\left[ \frac{1}{\Gamma
^{2}}\left( \frac{\sqrt{2k}}{1+e^{-2\sqrt{2k}}}-Q\right) ^{2}+1\right] ^{-1}}%
,  \label{A}
\end{equation}

\begin{equation}
B=\frac{1}{\Gamma }\left( \frac{\sqrt{2k}}{1+e^{-2\sqrt{2k}l}}-Q\right) A,
\label{BA}
\end{equation}%
\begin{equation}
Q\equiv \frac{\sqrt{2k}}{1-e^{-4\sqrt{2k}l}}\pm \sqrt{\frac{k}{2\sinh
^{2}\left( 2\sqrt{2k}l\right) }-\Gamma ^{2}}.  \label{QQ}
\end{equation}%
These solutions exist only in the region where the radical in Eq. (\ref{QQ})
is real, i.e.,%
\begin{equation}
\frac{\sqrt{2k}}{\sinh \left( 2\sqrt{2k}l\right) }\geq 2\Gamma .  \label{>}
\end{equation}%
Condition (\ref{>}) holds at $k\leq k_{\max }$, with $k_{\max }$ determined
by a transcendental equation,%
\begin{equation}
\frac{\sqrt{2k_{\max }}}{\sinh \left( 2\sqrt{2k_{\max }}l\right) }=2\Gamma .
\label{gamma}
\end{equation}%
It is easy to see that Eq. (\ref{gamma}) has a single physical solution
provided that 
\begin{equation}
\Gamma <\Gamma _{\mathrm{cr}}\equiv \left( 4l\right) ^{-1},  \label{1/4}
\end{equation}%
and no solutions at $\Gamma >\Gamma _{\mathrm{cr}}$, which is a
manifestation of the above-mentioned generic feature of nonlinear $\mathcal{%
PT}$-symmetric systems: soliton families exist below a certain critical
value of the gain-loss coefficient. On the other hand, Eq. (\ref{1/4})
demonstrates that the critical value diverges in the limit of $l\rightarrow
0 $, which corresponds to the replacement of the separated gain and loss by
the $\mathcal{PT}$ dipole in Eqs. (\ref{eq}) and (\ref{U}). The latter fact
helps to understand why the above analytical solutions, found in the $%
\mathcal{PT}$-dipole models, do not feature the existence threshold.

Finally, in the limit of $\Gamma \rightarrow \Gamma _{\mathrm{cr}}$, the
solution of Eq. (\ref{gamma}) is $k\rightarrow 0$, and Eqs. (\ref{A}), (\ref%
{BA}) and (\ref{QQ}) then yield $A=-B$, with 
\begin{equation}
A^{2}=A_{\mathrm{cr}}^{2}\equiv \left( 2\varepsilon _{2}\right) ^{-1}\left[
\left( 4l\right) ^{-1}-\varepsilon _{0}\right] ,  \label{Al}
\end{equation}%
while Eq. (\ref{CDAB}) yields $C=A$, $D=-A/\left( \sqrt{2k}l\right) $. Thus,
solution (\ref{in-out}) takes the eventual form%
\begin{eqnarray}
U_{\mathrm{out}}(x) &=&A_{\mathrm{cr}}\left[ 1-i~\mathrm{sgn}(x)\right] , 
\notag \\
U_{\mathrm{in}}(x) &=&A_{\mathrm{cr}}\left( 1-ix/l\right) ,  \label{lim}
\end{eqnarray}%
where $A_{\mathrm{cr}}$ is given by Eq. (\ref{Al}). Further, it is easy to
check directly that Eqs. (\ref{lim}) and (\ref{Al}) indeed give a particular
exact delocalized $\mathcal{PT}$-symmetric solution of Eq. (\ref{Gamma}),
provided that expression (\ref{Al}) yields $A_{\mathrm{cr}}^{2}>0$, i.e., $%
\varepsilon _{0}<\left( 4l\right) ^{-1}$ or $\varepsilon _{0}>\left(
4l\right) ^{-1}$ for $\varepsilon _{2}>0$ and $\varepsilon _{2}<0$,
respectively, cf. Eq. (\ref{><}).

\section{Numerical results for the regularized model}

Numerical results are presented below, chiefly, for solutions which are
counterparts of the analytical ones obtained above in the fully explicit
form, i.e., the solutions based on Eqs. (\ref{SF}), (\ref{SDF}), (\ref{theta}%
) and (\ref{xi}) (for the spatially uniform nonlinearity, with $\varepsilon
_{2}=0$) or (\ref{varepsilon_0=0}), for $\varepsilon _{0}=0$, i.e., the
purely nonlinear attractive potential at $x=0$.

\subsection{The approximation of the $\protect\delta $-function in numerical
solutions}

As said above, the numerical analysis of the model aims to obtain solutions
for the $\delta $-function replaced by its finite-width regularization, $%
\tilde{\delta}(x)$, with the objectives to produce solutions for a
modification of the model relevant for the experimental implementation, and
also to test the stability of the $\mathcal{PT}$-symmetric modes produced
above in the analytical form. In many works, $\tilde{\delta}(x)$ was used in
the form of a narrow Gaussian, see, e.g., Ref. \cite{Dong}. However, this is
not convenient in the present context, as, replacing the exact solutions in
the form of Eqs. (\ref{SF}), (\ref{SDF}), and (\ref{sigma=0}) by regularized
expressions, it is necessary, \textit{inter alia}, to replace $\mathrm{sgn}%
(x)\equiv -1+2\int_{-\infty }^{x}\delta (x^{\prime })dx^{\prime }$ by a
continuous function realized as $-1+2\int_{-\infty }^{x}\tilde{\delta}%
(x^{\prime })dx^{\prime }$, which would imply using a non-elementary
function in the case of the Gaussian. Therefore, we here use the
regularization in the form of the Lorentzian,%
\begin{equation}
\delta (x)\rightarrow \frac{a}{\pi }\frac{1}{x^{2}+a^{2}},~\delta ^{\prime
}(x)\rightarrow -\frac{2a}{\pi }\frac{x}{\left( x^{2}+a^{2}\right) ^{2}},~%
\mathrm{sgn}(x)\rightarrow \frac{2}{\pi }\arctan \left( \frac{x}{a}\right) ,
\label{Lorentz}
\end{equation}%
with $0<a\ll k^{-1/2}$.

\subsection{The self-focusing uniform nonlinearity ($\protect\sigma =+1,%
\protect\varepsilon _{2}=0$)}

We start the presentation of the results with the case of the uniform SF
nonlinearity, fixing $\varepsilon _{0}=1$ in Eqs. (\ref{eq}) and (\ref{U})
(larger values of $\varepsilon _{0}$ are used below to report results
obtained in the model with the uniform SDF \ nonlinearity). Stationary
solutions were found by solving Eq. (\ref{U}) with the help of the Newton's
method, using the input provided by the analytical solution in the form of
Eqs. (\ref{SF}), (\ref{theta}), and (\ref{xi}), with the regularization
implemented as per Eq. (\ref{Lorentz}). The stability of the so generated
solutions was tested through direct simulations of their perturbed evolution
by means of the fourth-order split-step method. It was implemented in domain 
$-10<x<+10$, with periodic boundary conditions (the width of the integration
domain is definitely much larger than the size of all the modes considered
in this work, see Figs. \ref{fig1}, \ref{fig2}, and \ref{fig7} below).
Sufficient numerical stability and accuracy were achieved with time and
space step sizes $\Delta t=0.001$ and $\Delta x=0.039$, respectively.
Accordingly, values $a\geq 0.02$ of the regularization scale were adopted in
Eq. (\ref{Lorentz}), as $a$ cannot be essentially smaller than $\Delta x$
(in fact, the plots displayed below are generated with $a=0.02$, unless it
is stated otherwise).

The first result is that, for fixed values of $a$ in Eq. (\ref{Lorentz}),
there is a critical value, $\gamma _{\mathrm{cr}}$, of the $\mathcal{PT}$
gain-loss coefficient, such that, at $\gamma <\gamma _{\mathrm{cr}}$, the
numerical solution features a shape very close to that of the analytical
solution corresponding to the ideal $\delta ^{\prime }$- and $\delta $-
functions, while at $\gamma > \gamma _{\mathrm{cr}}$ the single-peak shape
of the solution transforms into a \emph{double-peak} one, as shown in Fig. %
\ref{fig1}(a). In particular, it was found that $\gamma _{\mathrm{cr}}\left(
a=0.02\right) \approx 0.24$. As shown below, there is the second critical
value of $\gamma $ above which pinned modes do not exist at all, cf. the
exact result (\ref{1/4}). 
\begin{figure}[tbp]
\centering\subfigure[]{\includegraphics[width=3in]{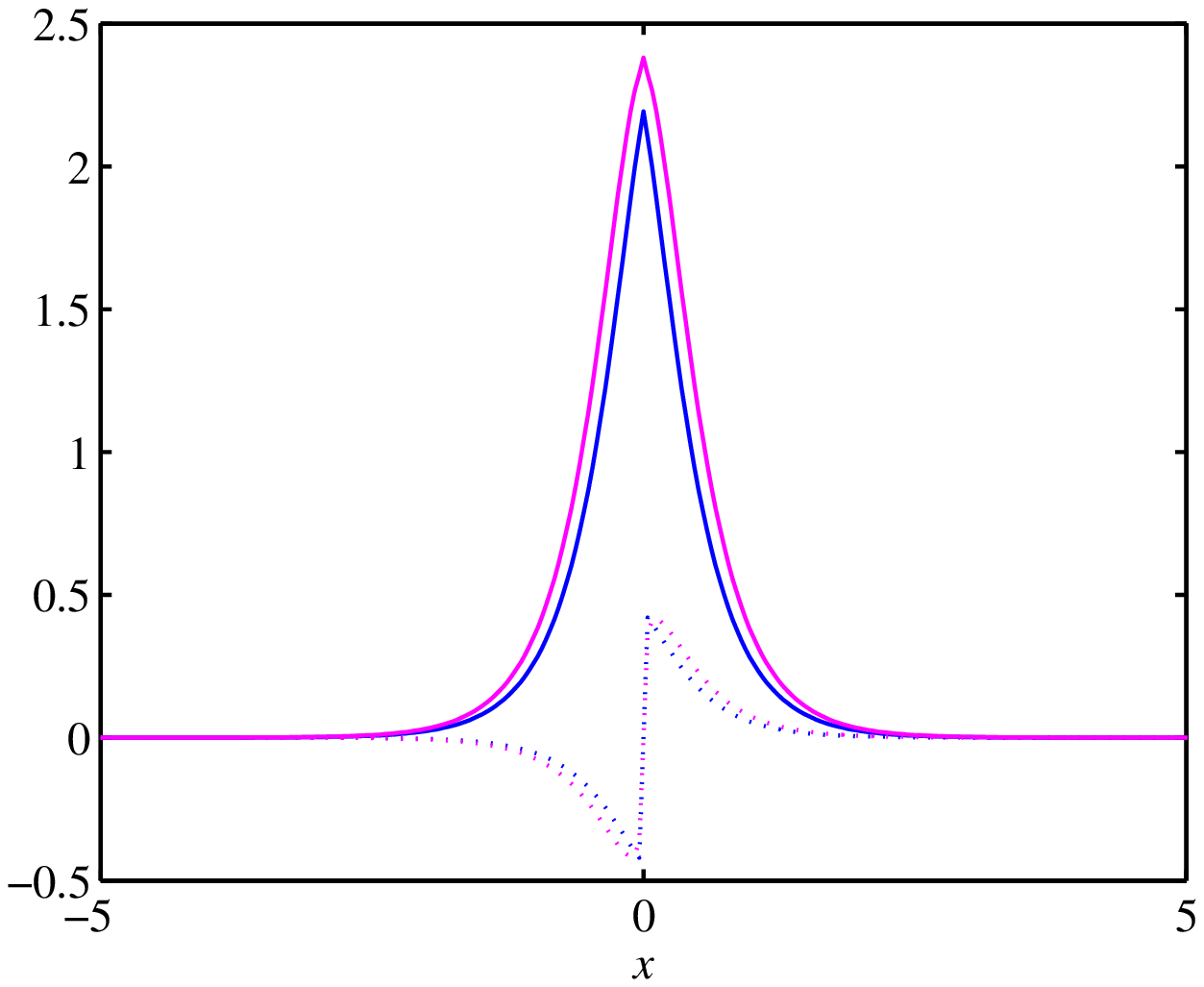}}%
\subfigure[]{\includegraphics[width=3in]{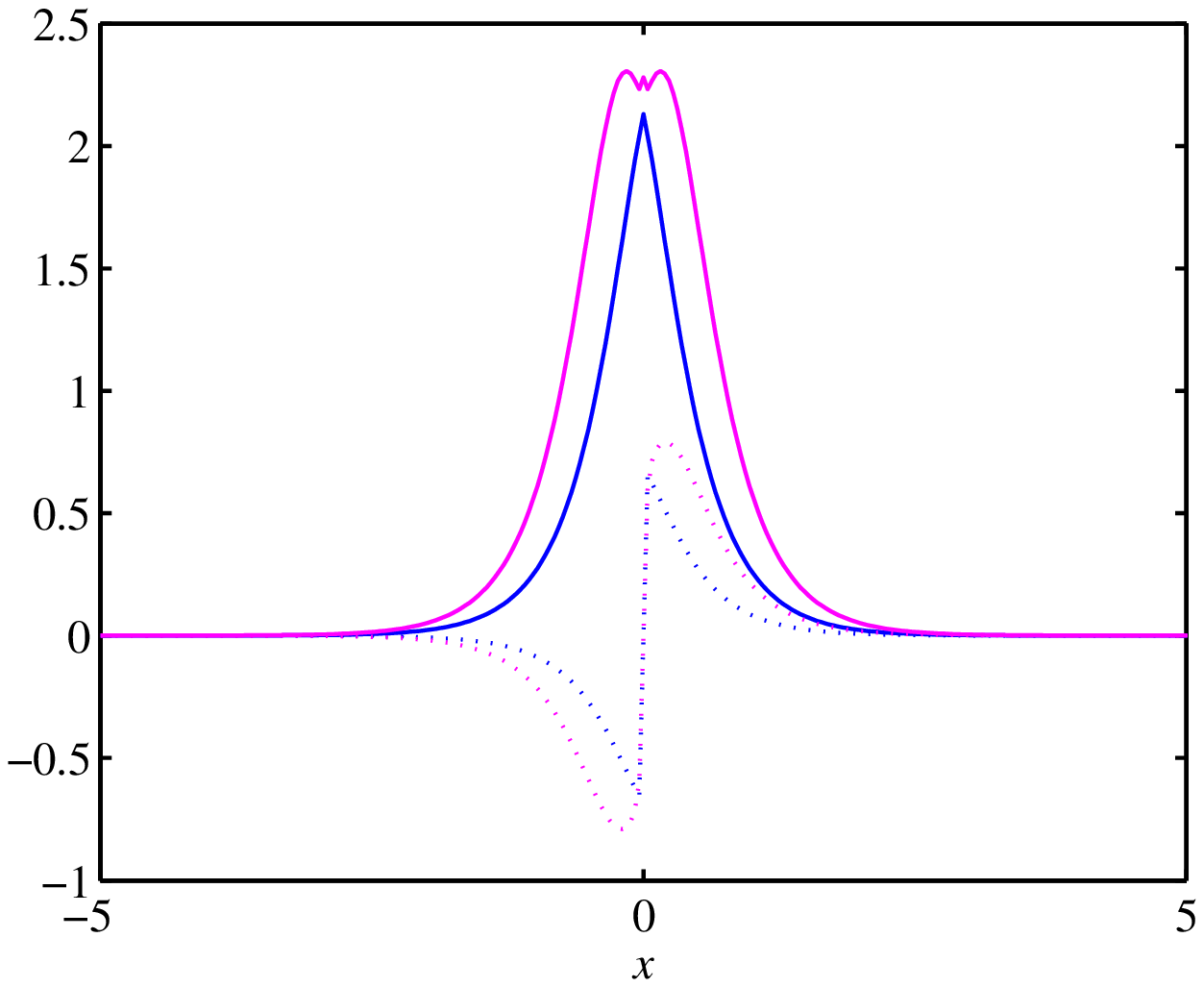}}
\caption{(Color online) Comparison between the analytical solutions (solid
and dotted blue curves show their real and imaginary parts, respectively),
given by Eqs. (\protect\ref{SF}), (\protect\ref{theta}), and (\protect\ref%
{xi}) with $\protect\sigma =+1,\protect\varepsilon _{2}=0,\protect%
\varepsilon _{0}=1$, and their numerically found counterparts, obtained by
means of regularization (\protect\ref{Lorentz}) with $a=0.02$ (magenta
curves). The $\mathcal{PT}$ gain-dissipation parameter is $\protect\gamma %
=0.20$ in (a) and $0.32$ in (b). In both panels, the solutions are produced
for propagation constant $k=3$.}
\label{fig1}
\end{figure}

The drastic difference between the single- and double-peak modes is that the
former ones are completely stable, as confirmed by systematic simulations
(not shown here in detail), while all the double-peak solutions are
unstable. This correlation between the shape and (in)stability of the pinned
modes is not surprising: the single- and double-peak structures imply that
the pinned mode is feeling, respectively, effective attraction to or
repulsion from the local defect. Accordingly, in the latter case the pinned
soliton is unstable against spontaneous detachment (escape) from the $%
\mathcal{PT}$ dipole, transforming itself into an ordinary freely moving NLS
soliton, see an example in Fig. \ref{fig2}. 
\begin{figure}[tbp]
\centering\includegraphics[width=4in]{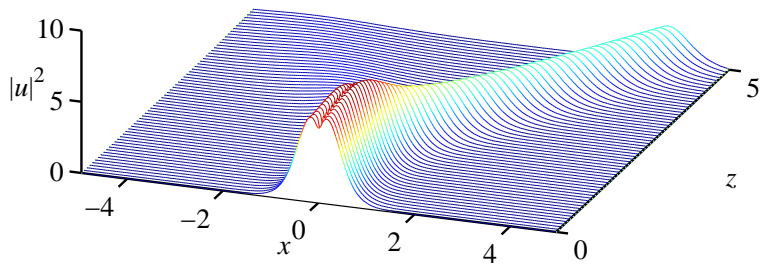}
\caption{(Color online) The unstable evolution (escape) of the double-peak
soliton from Fig. \protect\ref{fig1}(b).}
\label{fig2}
\end{figure}

The transition between stable single-peak and unstable double-peak pinned
modes was earlier reported in Ref. \cite{Barash}, which was dealing with a
chain of parametrically driven damped pendulums, with a discrete soliton
attached to a local defect in the chain. In that case, the instability
development was different, leading not to detachment of the soliton, but
rather to $\pi $-out-of-phase oscillations of the two lobes of the
double-peak structure.

\begin{figure}[tbp]
\centering\subfigure[]{\includegraphics[width=3in]{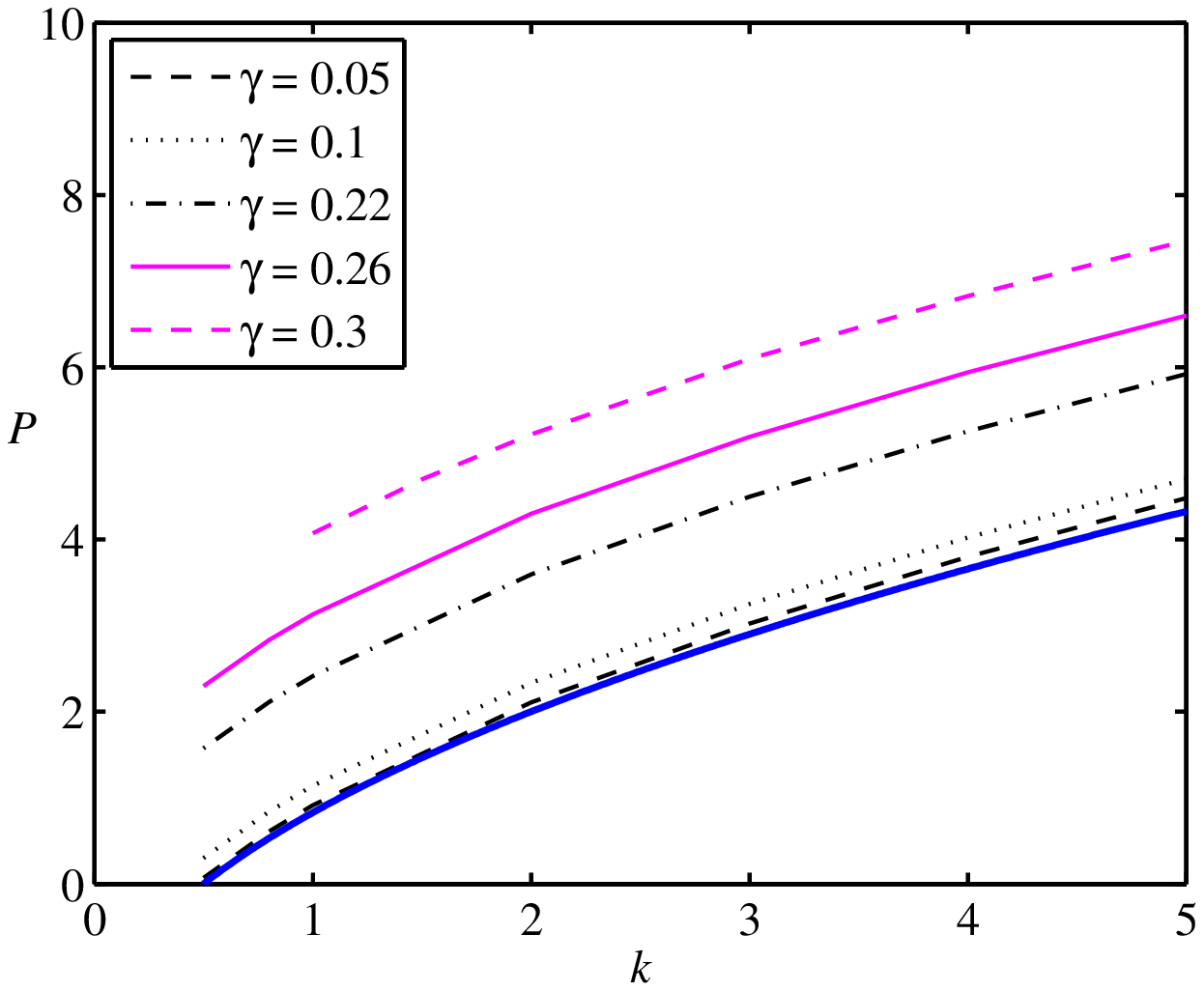}}%
\subfigure[]{\includegraphics[width=3in]{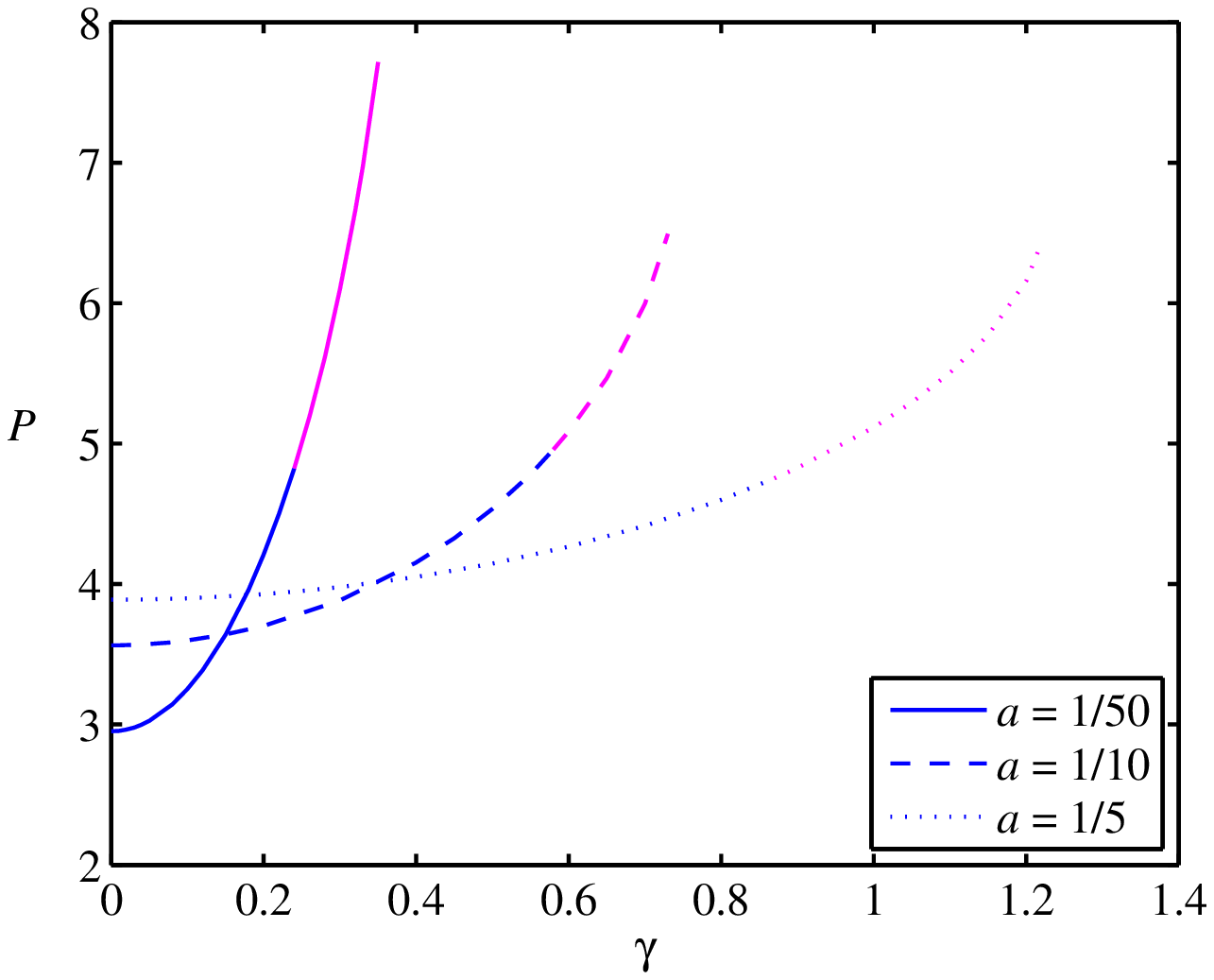}}
\caption{(Color online) (a) Total power $P$ versus propagation constant $k$
at fixed values of the $\mathcal{PT}$ gain-loss coefficient, $\protect\gamma 
$, in the system with $\protect\sigma =+1,\protect\varepsilon _{2}=0,\protect%
\varepsilon _{0}=1$, and regularization scale $a=0.02$ in Eq. (\protect\ref%
{Lorentz}). The blue line (the bottom one) shows the analytical result (%
\protect\ref{Power}), while the black and magenta lines (three intermediate
and two top ones, respectively) represent, respectively, the numerically
found stable (single-peak) and unstable (double-peak) modes. (b) $P(\protect%
\gamma )$ for fixed $k=3.0$ and different fixed values of $a$. Blue and
magenta segments of the curves (bottom and top ones, respectively) represent
the single- and double-peak (stable and unstable) pinned solitons,
respectively.}
\label{fig3}
\end{figure}

The results for soliton families in the present situation are summarized in
Figs. \ref{fig3}(a) and \ref{fig3}(b), in the form of plots for $P(k)$ [cf.
Eq. (\ref{Power})] and $P(\gamma )$ . The plots also delineate the effective
boundary between the stable single-peak modes and unstable double-peak ones.
The curves in panel \ref{fig3}(b) terminate at critical points, beyond which
no pinned modes are produced by the numerical solution. Exact analytical
results for solitons in the model of the $\mathcal{PT}$-symmetric nonlinear
coupler \cite{dual} suggest that the termination of the solution branches
may be explained by a tangent (saddle-node) bifurcation, i.e., annihilation
of the given branch with an additional unstable one (in the coupler model,
this is the branch of $\mathcal{PT}$-antisymmetric solitons). However,
search for that additional branch in the present model, which is,
presumably, a fully unstable one, is a challenging problem. 

Note that, at $\gamma \lesssim 0.1$, the numerically found total power
almost does not depend on $\gamma $ in Fig. \ref{fig3}(a), in accordance
with analytical result (\ref{Power}). However, $P$ grows with $\gamma $ at
larger values of $\gamma $. The analytical curve in Fig. \ref{fig3}(a)
terminates at $k=0.5$, as predicted by Eq. (\ref{xi}) for $\varepsilon
_{0}=1 $, but with the growth of $\gamma $ the cutoff value of $k$ increases.

\begin{figure}[tbp]
\centering\includegraphics[width=3in]{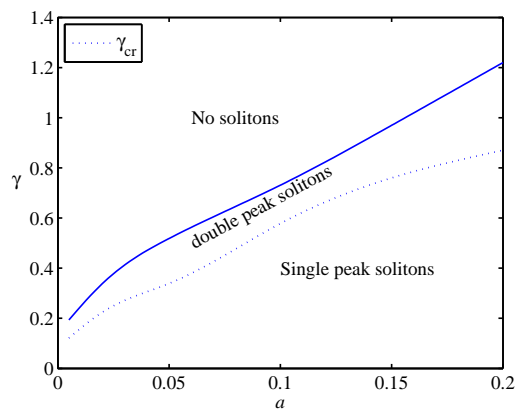}
\caption{(Color online) Regions of the existence of stable single-peak and
unstable double-peak solitons, separated by $\protect\gamma =\protect\gamma %
_{\mathrm{cr}}(a)$, in the plane of the regularization scale, $a$, and $%
\mathcal{PT}$ gain-loss parameter, $\protect\gamma ,$ for fixed $k=3.0$, in
the system with $\protect\sigma =+1,\protect\varepsilon _{2}=0,\protect%
\varepsilon _{0}=1$. }
\label{fig4}
\end{figure}

Finally, Fig. \ref{fig4} summarizes the findings in the plane of $\left(
a,\gamma \right) $ for fixed $k=3.0$. It is clearly seen that the region of
the unstable double-peak solitons is actually a relatively narrow boundary
layer between the broad areas in which the stable single-peak solitons
exist, or no solitons exist at all, at large values of $\gamma $. Note also
that the stability area strongly expands to larger values of $\gamma $ as
the regularized profile (\ref{Lorentz}) becomes smoother, with the increase
of $a$. On the other hand, the stability region remains finite even for very
small $a$. 

\subsection{The system with the self-focusing uniform nonlinearity and
nonlinear pinning potential ($\protect\sigma =+1,\protect\varepsilon %
_{2}\neq 0$)}

\begin{figure}[tbp]
\centering\subfigure[]{\includegraphics[width=3in]{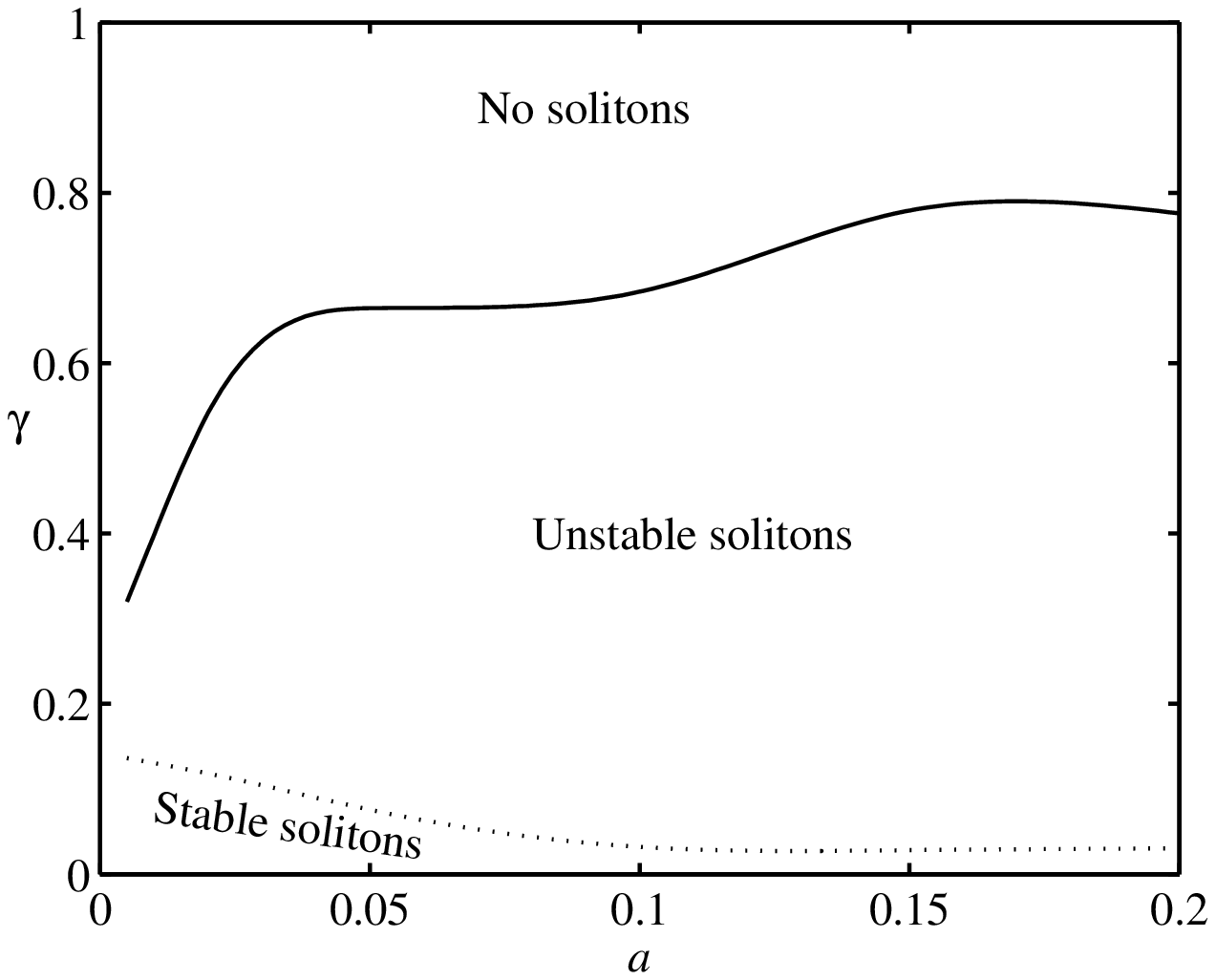}}%
\subfigure[]{\includegraphics[width=3in]{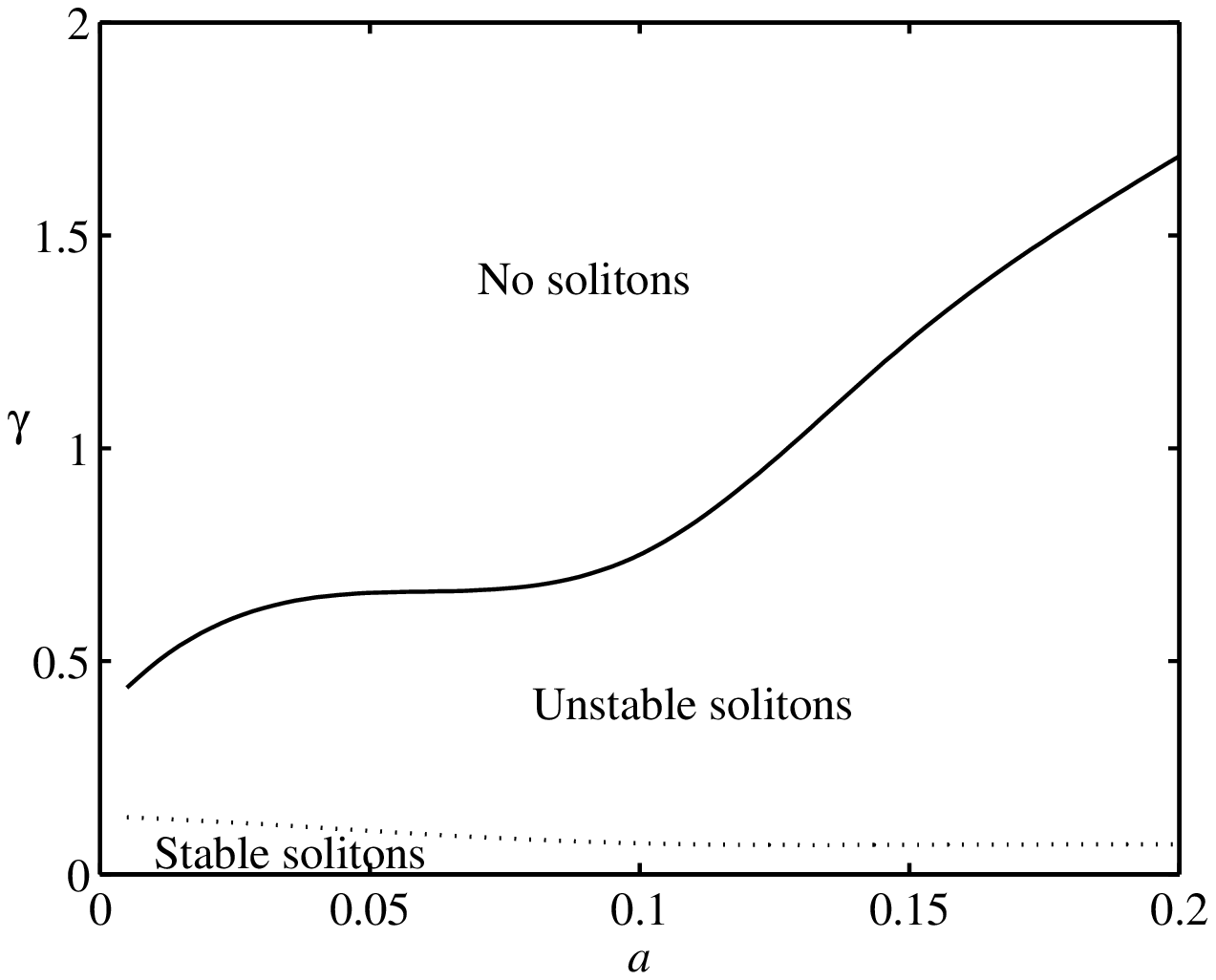}}
\caption{The same as in Fig. \protect\ref{fig4} (for $\protect\sigma %
=+1,k=3.0$), but in the model with the purely nonlinear attractive
potential, i.e., $\protect\varepsilon _{0}=0$, and $\protect\varepsilon %
_{2}=2.0$ (a) or $\protect\varepsilon _{2}=8.0$ (b). As before, in the
present case stable and unstable solitons feature the single- and
double-peak shapes, respectively.}
\label{fig5}
\end{figure}

Another explicit solution produced above, based on Eqs. (\ref{SF}), (\ref%
{theta}), and (\ref{varepsilon_0=0}), pertains to the case of $\sigma =+1$, $%
\varepsilon _{0}=0$, and $\varepsilon _{2}>0$, when the attractive potential
of the $\mathcal{PT}$ dipole is purely nonlinear. In this case, stable
single-peak modes, close to the aforementioned analytical solution, were
found for $0\leqslant \gamma \leqslant 0.13$, while at $\gamma >0.13$ the
pinned modes are unstable, featuring a double-peak shape. Similar to Fig. %
\ref{fig4}, the existence and stability diagram for the solitons is plotted
in the plane of $\left( a,\gamma \right) $ in Fig. \ref{fig5} for smaller
(a) and larger (b) values of $\varepsilon _{2}$. The comparison with Fig. %
\ref{fig4} demonstrates that, in the case of the nonlinear pinning
potential, the stability area is much smaller than it was in the case of the
linear attractive potential. 

\begin{figure}[tbp]
\centering\subfigure[]{\includegraphics[width=3in]{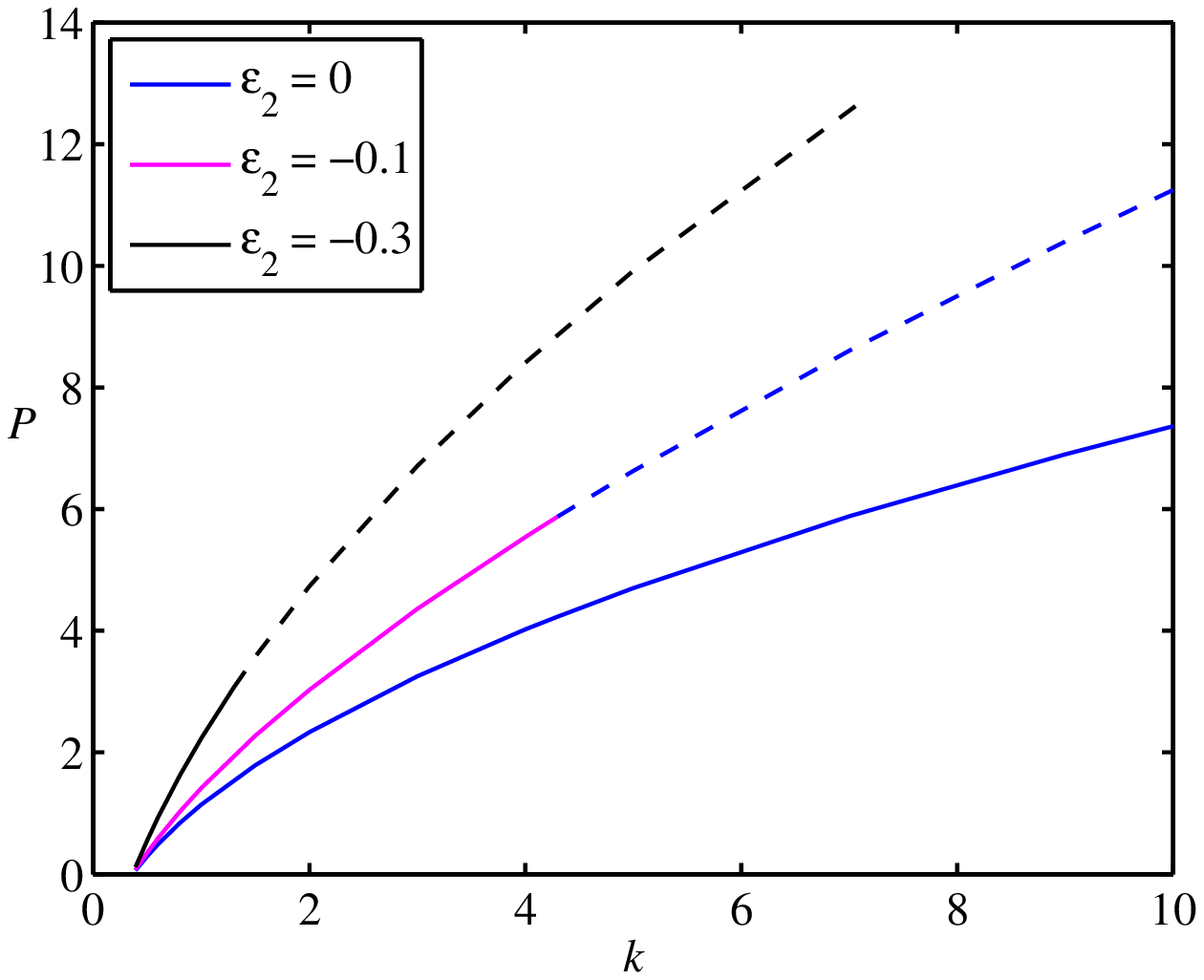}}%
\subfigure[]{\includegraphics[width=3in]{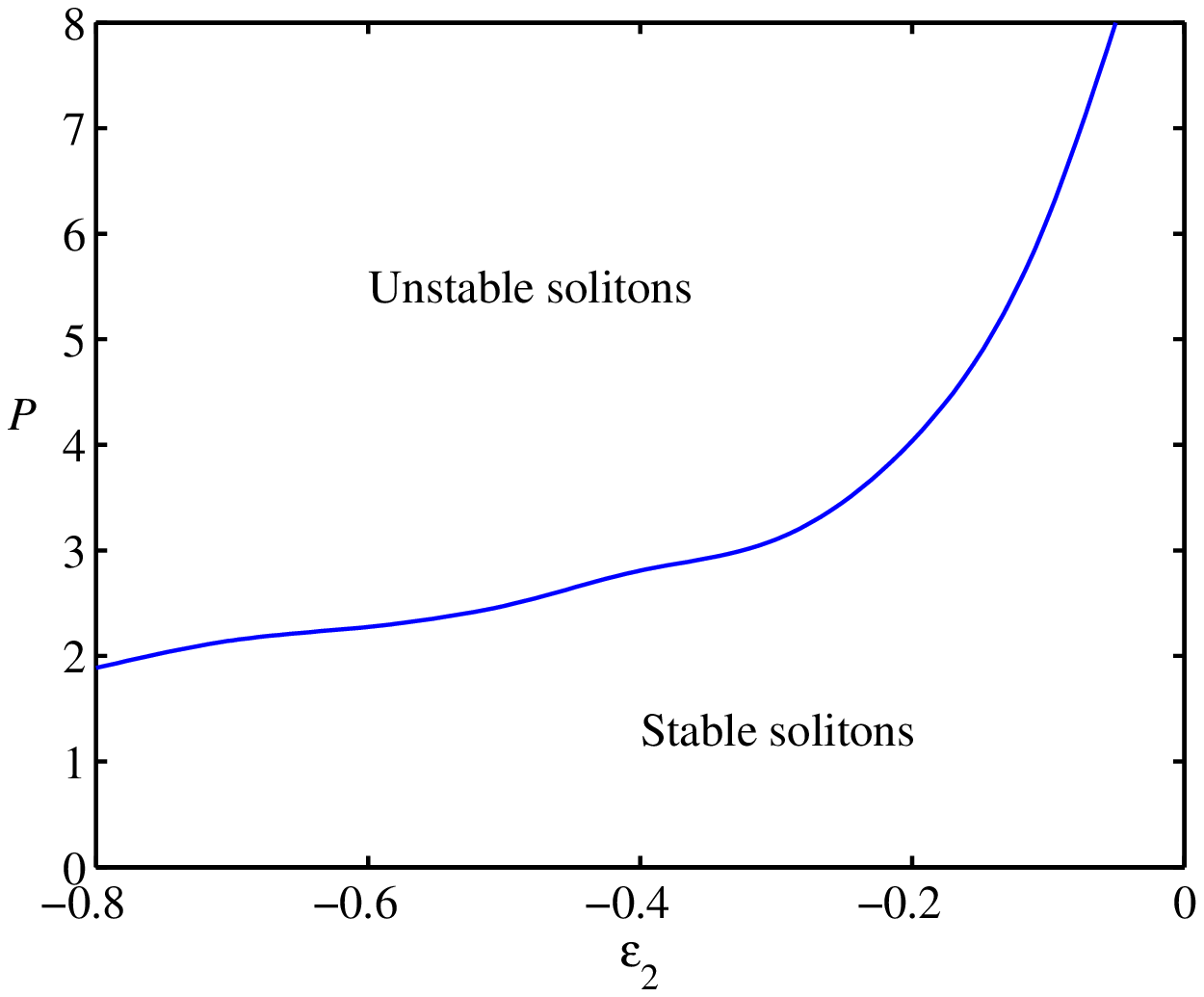}}
\caption{(Color online) (a) $P(k)$ curves in the model with $\protect\sigma %
=+1$, $\protect\varepsilon =1$, $\protect\gamma =0.1$ and different values
of $\protect\varepsilon _{2}\leq 0$ (the increase of $\protect\varepsilon %
_{2}$ from $-0.3$ to $0$ corresponds to the transition from the top curve to
the bottom one). The solid and dashed segments represent stable single-peak
solitons and unstable double-peak ones, respectively. (b) The stability
boundary in the plane of $\left( \protect\varepsilon _{2},P\right) $ for the
same case.}
\label{fig6}
\end{figure}

Next, we consider combined linear and nonlinear pinning potentials,
restoring $\varepsilon _{0}=1$. With $\varepsilon _{2}>0$, stable
single-peak solitons are readily found up to the respective critical value
of $\gamma $ (for instance, at $k=2.0$ they are found at $\gamma <0.15$ for
any value of $\varepsilon _{2}$). With $\varepsilon _{2}<0$, both stable
single-peak solitons and unstable double-peak ones are produced by the
numerical solution. For this case, Fig. \ref{fig6}(a) displays $P(k)$ curves
with fixed $\gamma =0.1$ and different negative values of $\varepsilon _{2}$%
. The curves include segments representing both the single- and double-peak
modes. Further, the respective stability boundary in the plane of $\left(
\varepsilon _{2},P\right) $ for fixed $\gamma =0.1$ is shown in Fig. \ref%
{fig6}(b). The increase of $P$ naturally leads to destabilization of the
pinned mode, as the corresponding nonlinear repulsive potential, accounted
for by $\varepsilon _{2}<0$, becomes stronger. 

\subsection{The self-defocusing nonlinearity ($\protect\sigma =-1$)}

\begin{figure}[tbp]
\centering\subfigure[]{\includegraphics[width=3in]{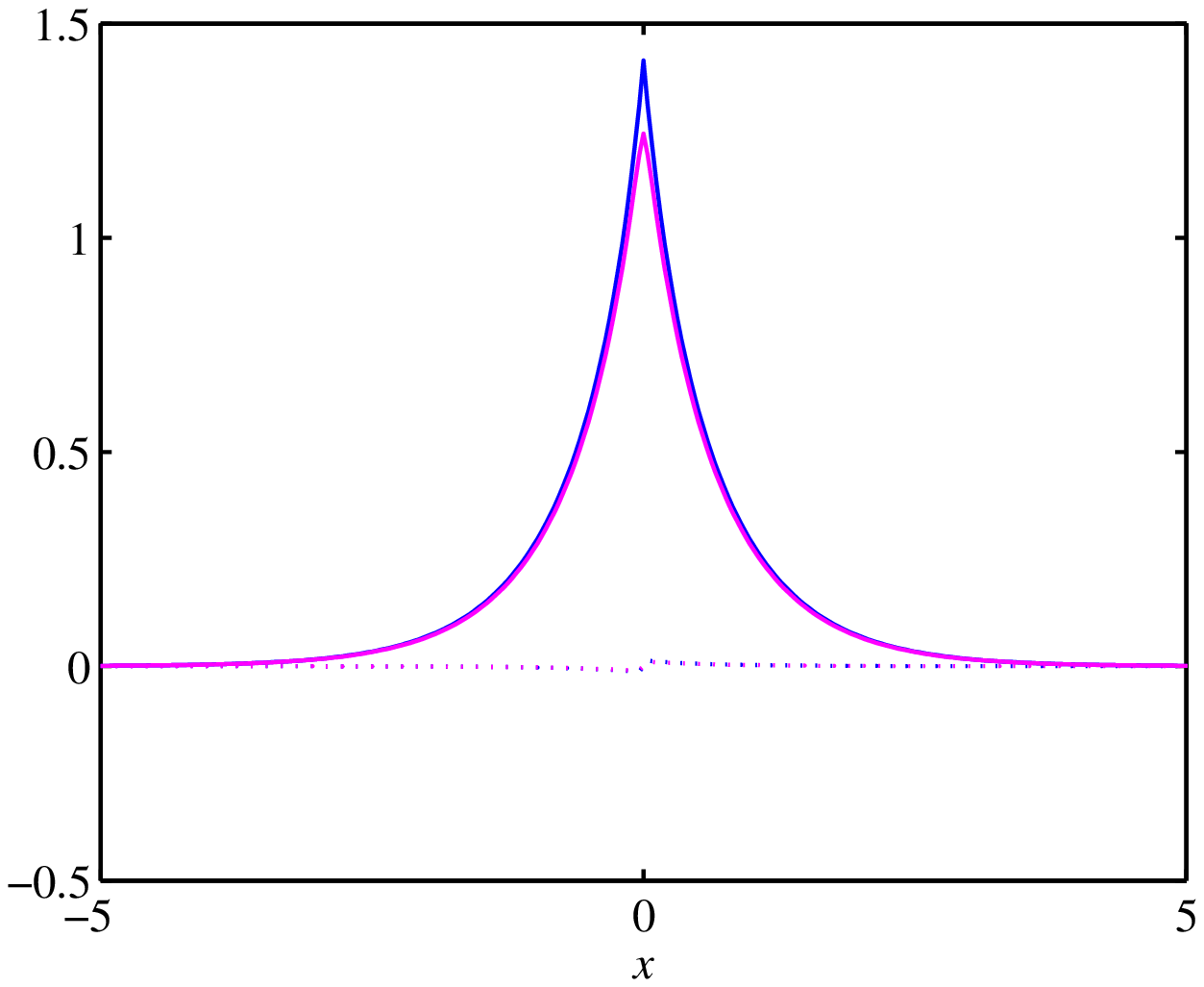}}%
\subfigure[]{\includegraphics[width=3in]{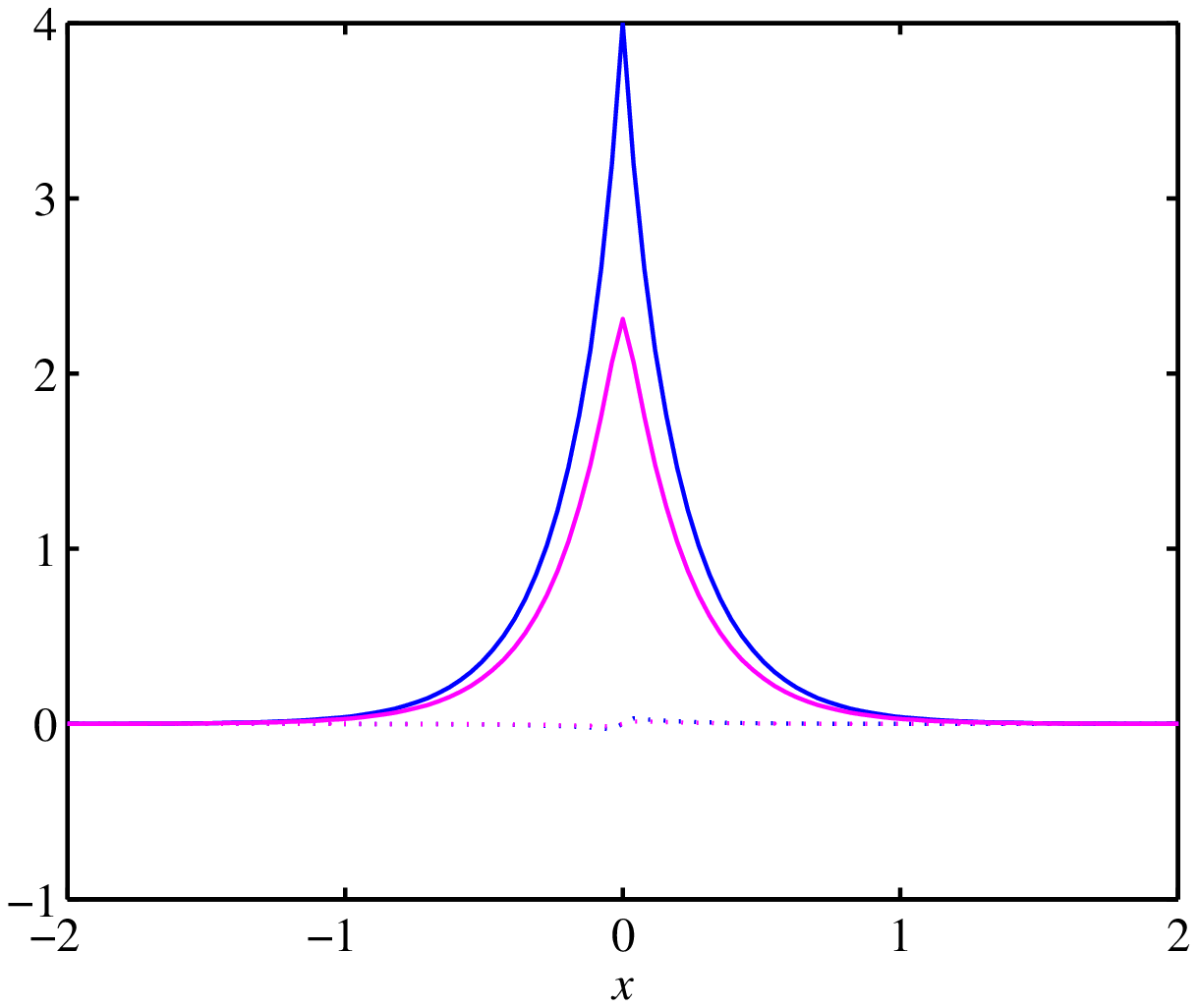}}
\caption{(Color online) Comparison between the analytical solutions (solid
and dotted blue curves show its real and imaginary parts, respectively),
given by Eqs. (\protect\ref{SDF}), (\protect\ref{theta}), and (\protect\ref%
{xi}) with $\protect\sigma =-1,\protect\varepsilon _{2}=0$, and their
numerically found counterparts, obtained by means of regularization (\protect
\ref{Lorentz}) with $a=0.02$ (lower magenta curves). Other parameters are $%
\protect\varepsilon _{0}=2.0$, $k=1.0$ in (a), and $\protect\varepsilon %
_{0}=6.0$, $k=10.0$ in (b). }
\label{fig7}
\end{figure}

\begin{figure}[tbp]
\centering\subfigure[]{\includegraphics[width=3in]{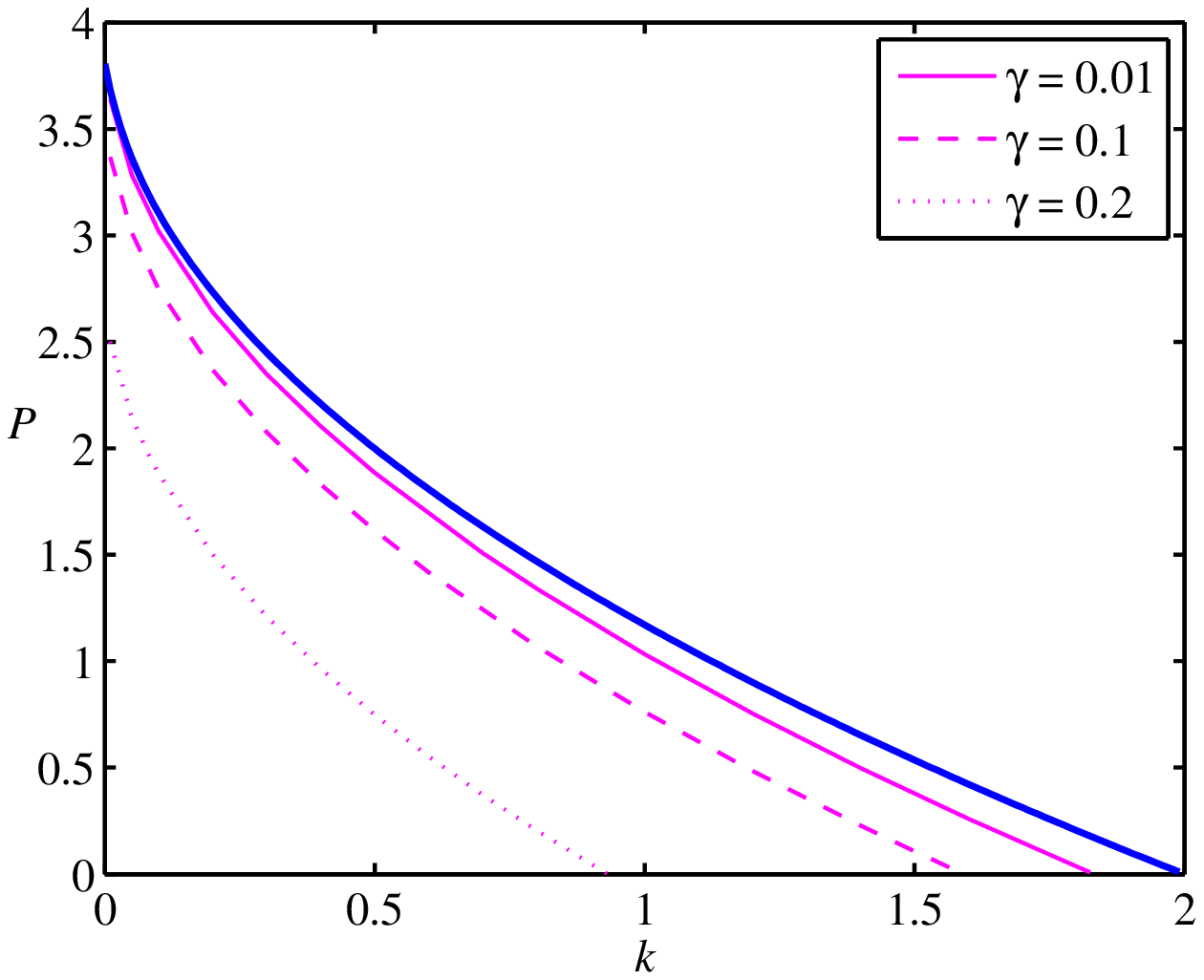}}%
\subfigure[]{\includegraphics[width=3in]{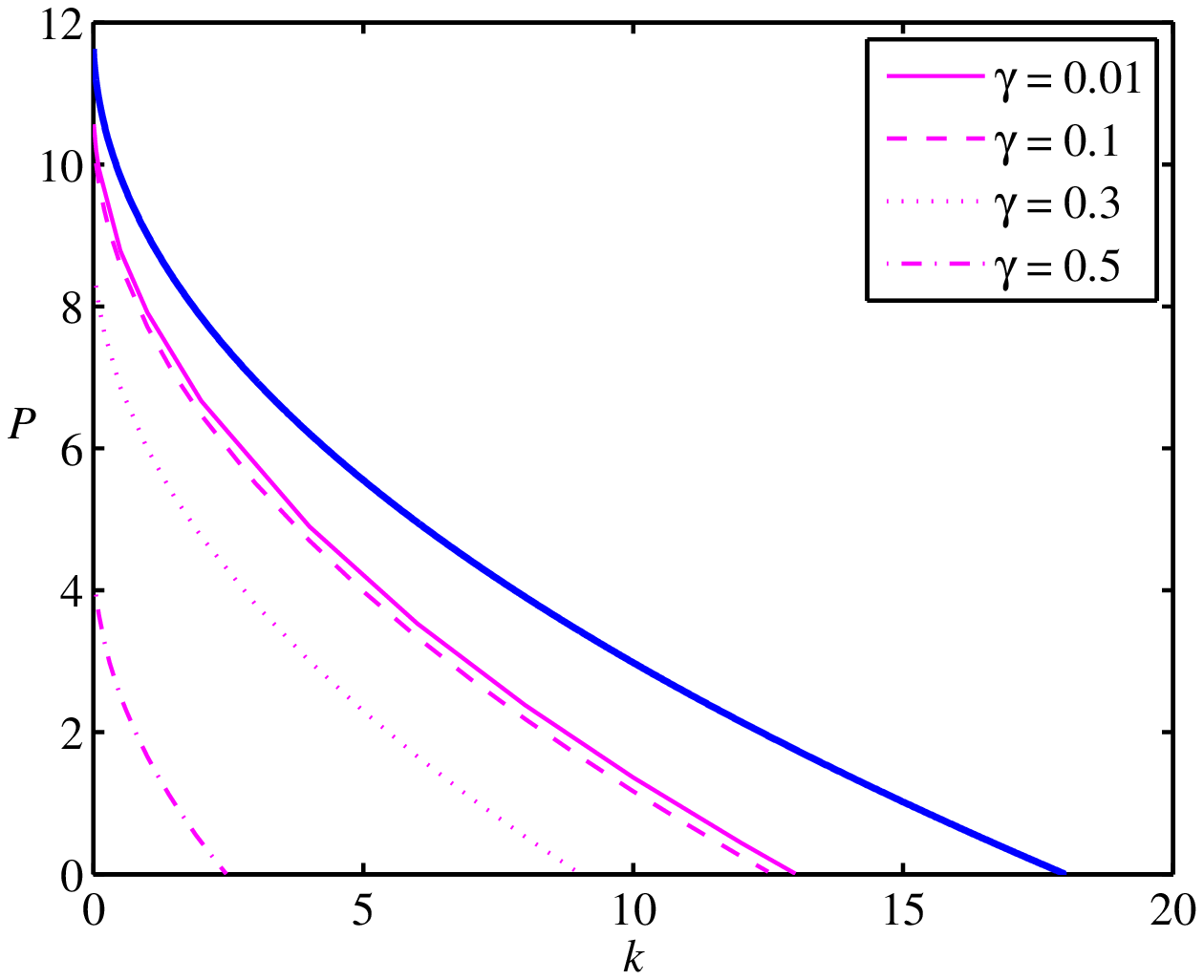}}
\caption{(Color online) (a) Curves $P(k)$ in the model with the uniform SDF
nonlinearity, i.e., $\protect\sigma =-1,\protect\varepsilon _{2}=0$, for
different values of $\protect\gamma $ and $\protect\varepsilon _{0}=2.0$ (a)
or $\protect\varepsilon _{0}=6.0$ (b). The blue (top) curves represent the
respective analytical result (\protect\ref{Power}), while the curves
generated by the numerical solution are plotted in magenta.}
\label{fig8}
\end{figure}

Another basic case corresponds to the exact solutions for the SDF
nonlinearity, given by Eqs. (\ref{SDF}), (\ref{theta}), and \ref{xi}) with $%
\sigma =-1,\varepsilon _{2}=0$. In this case, the numerical solution reveals
solely single-peak modes. For small $\gamma =0.01$, the comparison between
the analytical solutions and their numerically found counterparts in
displayed in Fig. \ref{fig7}. Further, soliton families are represented by
the respective $P(k)$ curves in Fig. \ref{fig8}. The corresponding
analytical dependence, given by Eq. (\ref{Power}), does not depend on $%
\gamma $, while its numerical counterparts deviate from it with the increase
of $\gamma $, especially for larger $\varepsilon _{0}$, see Fig. \ref{fig8}%
(b). 

Note that, in the framework of Eq. (\ref{U}), different values of $%
\varepsilon _{0}$ can be transformed into $\varepsilon _{0}=1$ by rescaling,
but regularization (\ref{Lorentz}) then implies that $a$ will be rescaled by
factor $\varepsilon _{0}$, hence larger $\varepsilon $ implies a farther
departure from the model with the ideal $\delta ^{\prime }$- and $\delta $-
functions. This fact explains the stronger deviation of the numerically
found curves from their analytically obtained counterparts in Fig. \ref{fig8}%
(b) in comparison with \ref{fig8}(a). The same trend is observed below in
Fig. \ref{fig9}.

In the model with the uniform SDF nonlinearity, the analytical solution
exists in the interval of the propagation constant $k<k_{\max }=\varepsilon
_{0}^{2}/2$, see Eq. (\ref{><}). The respective numerically found existence
boundaries for the single-peak solitons are displayed in Fig. \ref{fig9}.
The observed deviation of the boundary value from $k_{\max }$ at $\gamma =0$
is explained by the difference of the regularized $\delta $-function (\ref%
{Lorentz}) from its ideal counterpart, the existence region further
shrinking with the increase of $\gamma $. 
\begin{figure}[tbp]
\centering\subfigure[]{\includegraphics[width=3in]{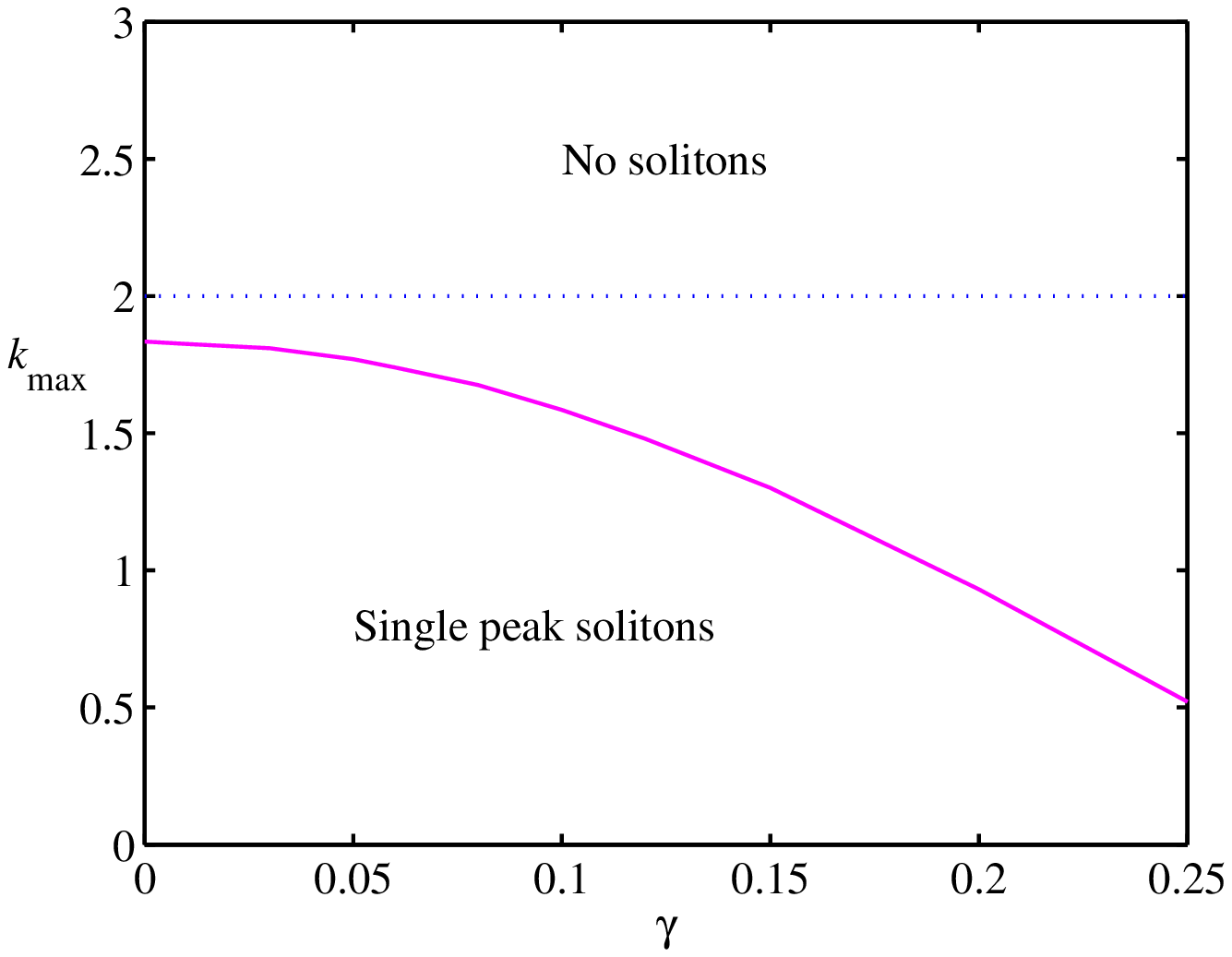}}%
\subfigure[]{\includegraphics[width=3in]{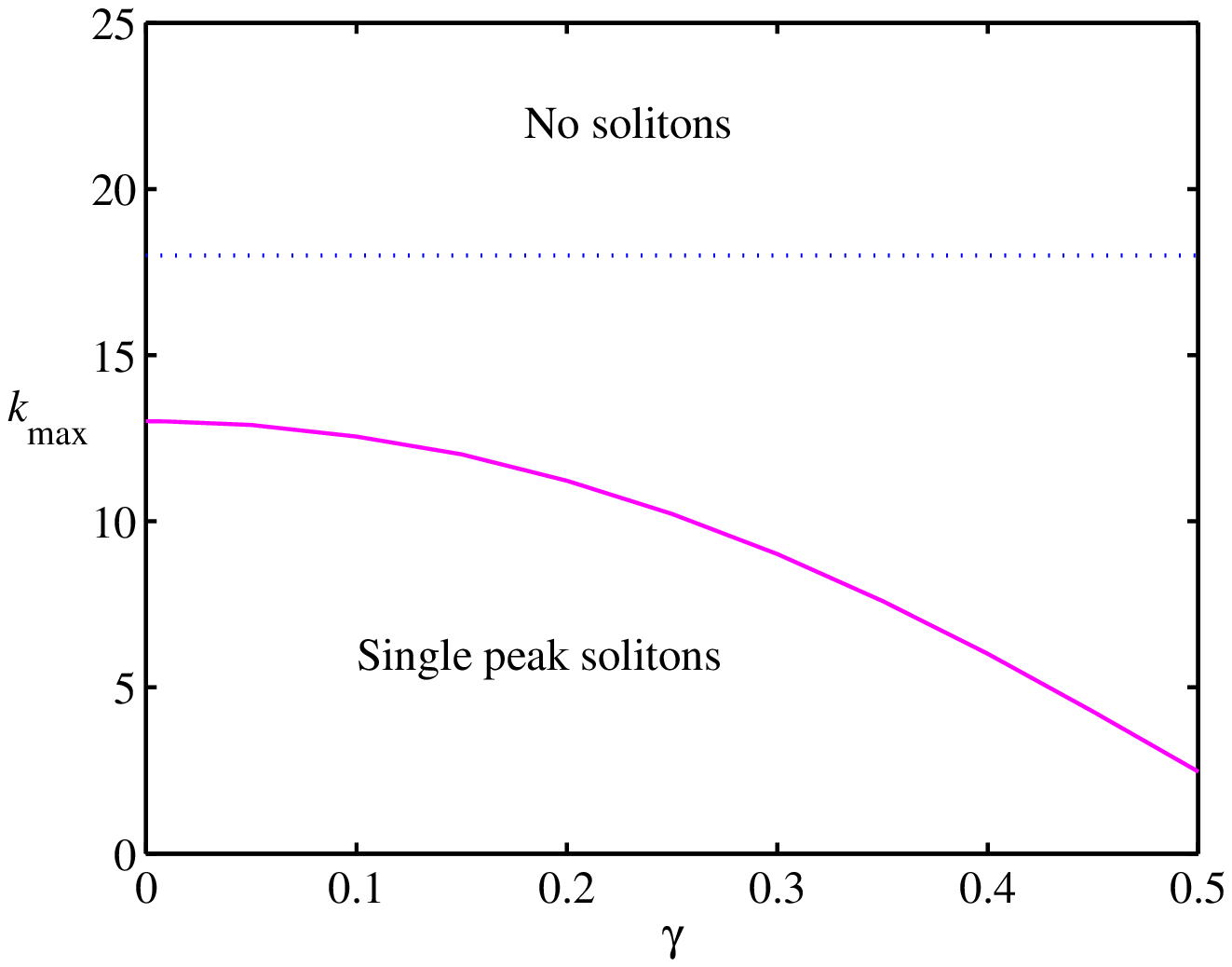}}
\caption{(Color online) The existence region for stable single-peak solitons
in the plane of $\left( \protect\gamma ,k\right) $ for the model with $%
\protect\sigma =-1$, $\protect\varepsilon _{2}=0$, and $\protect\varepsilon %
_{0}=2.0$ (a), or $\protect\varepsilon _{0}=6.0$ (b). The blue dotted
horizontal lines correspond to $k_{\max }=\protect\varepsilon _{0}^{2}/2$
predicted by analytical solution (\protect\ref{><}).}
\label{fig9}
\end{figure}

\begin{figure}[tbp]
\centering\includegraphics[width=3in]{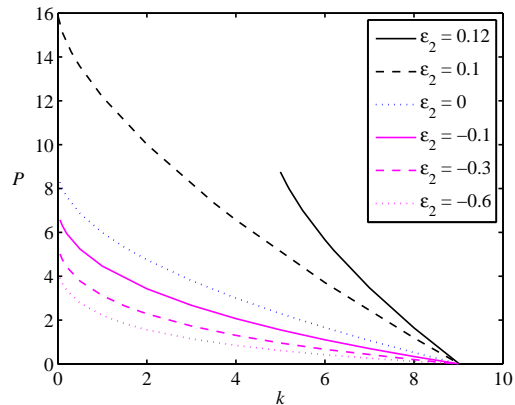}
\caption{(Color online) $P(k)$ curves in the model with $\protect\sigma =-1$%
, $\protect\varepsilon _{0}=6.0$, $\protect\gamma =0.3$ and different
positive and negative values of $\protect\varepsilon _{2}$. The branch for $%
\protect\varepsilon =0.12$ terminates at a point past which pinned modes
could not be found.}
\label{fig10}
\end{figure}

The results were extended to the case of $\varepsilon _{2}\neq 0$, when the
the pinning potential contains the nonlinear attractive or repulsive part,
corresponding to $\varepsilon _{2}>0$ and $\varepsilon _{2}<0$,
respectively. Stable single-peak solitons were found for either sign of $%
\varepsilon _{2}$. They are represented by the corresponding $P(k)$ curves
in Fig. \ref{fig10} for fixed $\varepsilon _{0}=6.0$ and $\gamma =0.3$,
which originate, at $P=0$, from point $k\approx 9$, in accordance with what
Fig. \ref{fig9}(b) shows for $\gamma =0.3$. 

Finally, the numerical results demonstrate that, in the case of $\sigma =-1$
and $\varepsilon _{0}=0$, solitons existing under the action of the
nonlinear pinning potential with $\varepsilon _{2}>0$, as predicted by Eqs. (%
\ref{SDF}), (\ref{theta}) and (\ref{varepsilon_0=0}), are completely
unstable. Recall that, unlike this result, in the model with the SF bulk
nonlinearity ($\sigma =+1$), a small stability area was found for solitons
pinned by the nonlinear potential, see Fig. \ref{fig5}.

\subsection{The linear host medium ($\protect\sigma =0$)}

\begin{figure}[tbp]
\centering\subfigure[]{\includegraphics[width=3in]{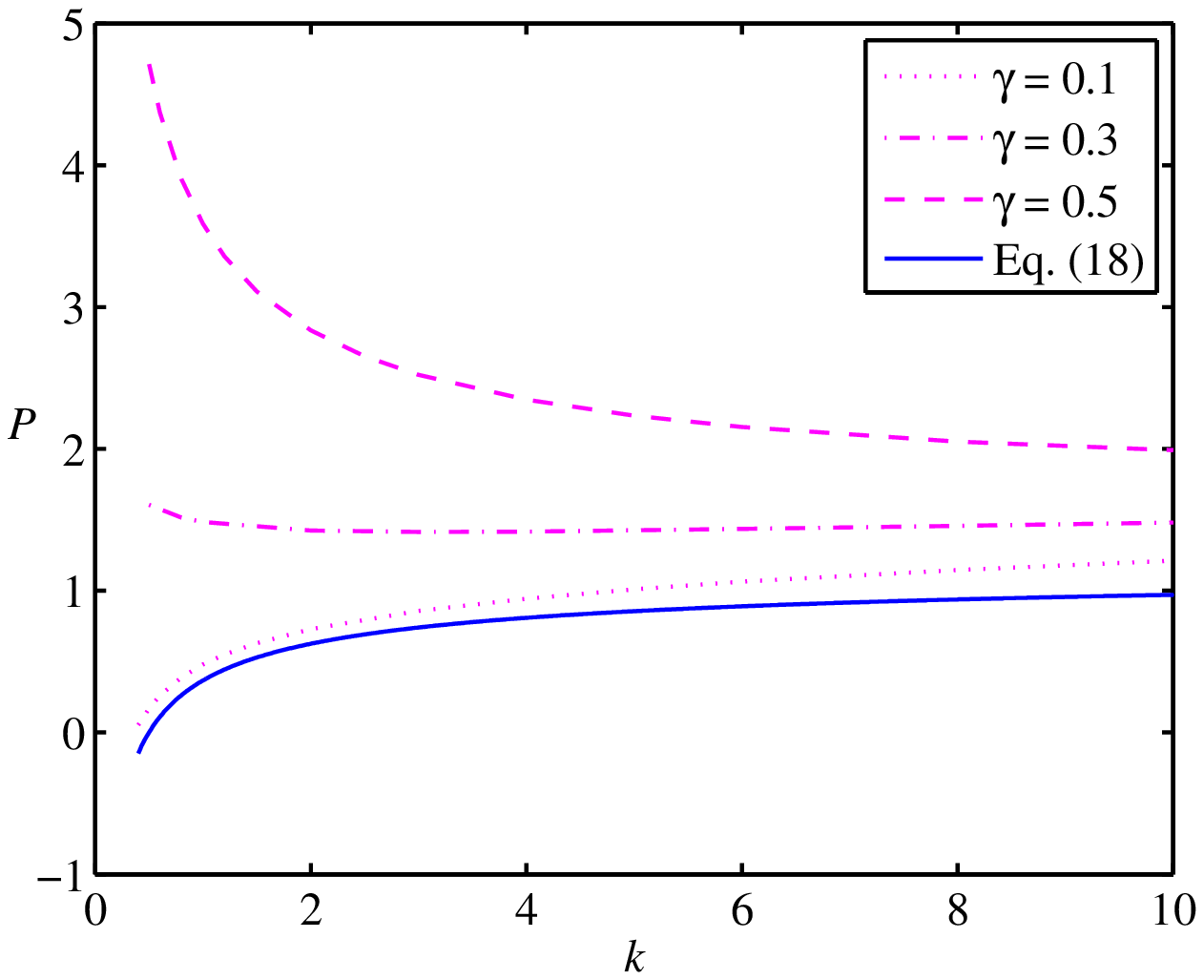}}%
\subfigure[]{\includegraphics[width=3in]{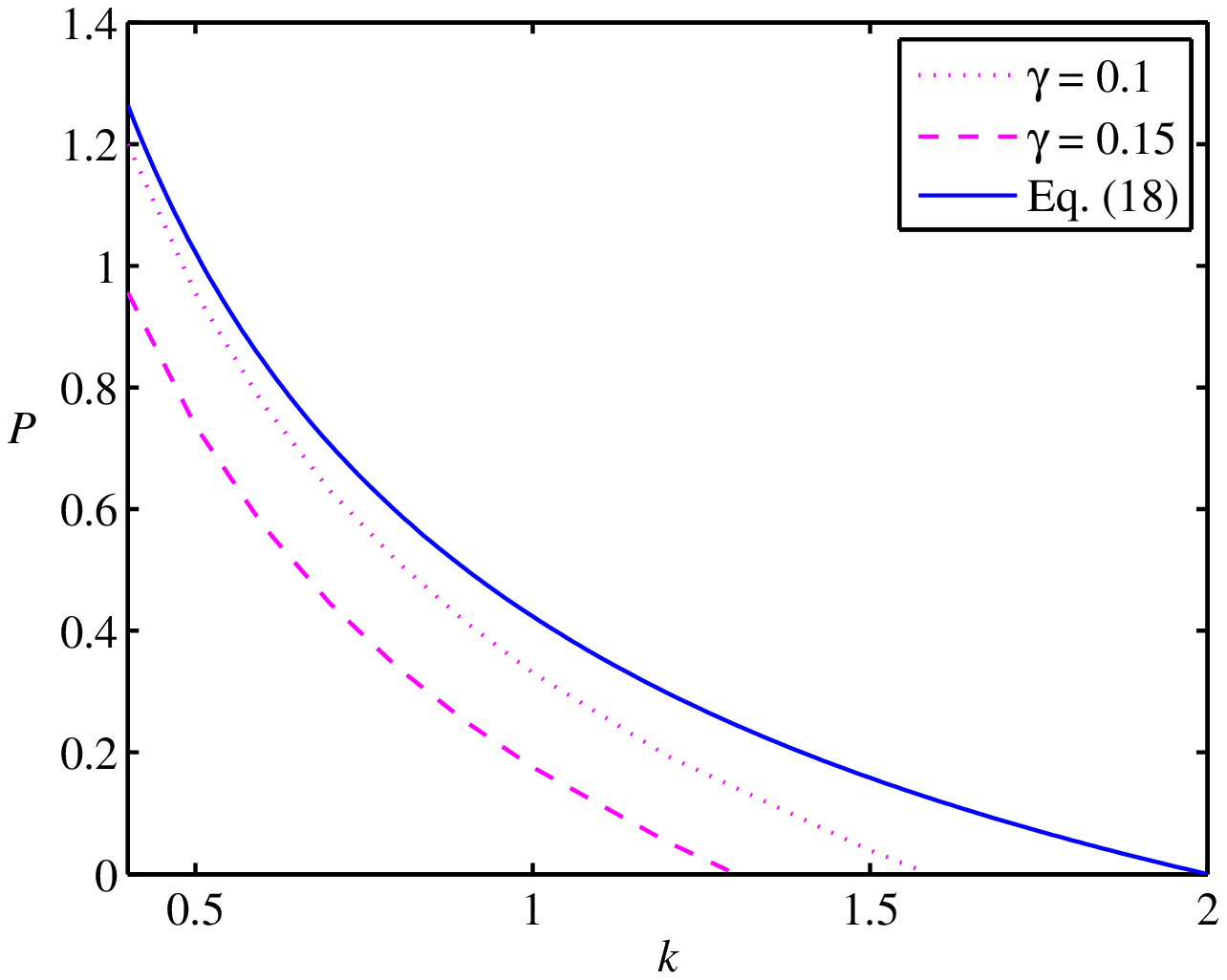}}
\caption{(Color online) $P(k)$ curves for pinned modes in the model with the
linear host medium, $\protect\sigma =0$. (a) The system with $\protect%
\varepsilon _{0}=1.0$, $\protect\varepsilon _{2}=+1$ and different values of 
$\protect\gamma $. The continous blue curve depicts the analytical result
given by Eq. (\protect\ref{P0}). (b) The same, but with $\protect\varepsilon %
_{0}=2.0$, $\protect\varepsilon _{2}=-1$. The numerical results displayed
here are obtained using regularization (\protect\ref{Lorentz}) with $a=0.02$%
. }
\label{fig11}
\end{figure}

Finally, we have produced numerical counterparts of the simplest exact
solutions given by Eq. (\ref{sigma=0}) for the nonlinear dipole embedded
into the linear medium ($\sigma =0,\varepsilon _{2}=\pm 1$). The results are
summarized in the form of $P(k)$ curves, which are displayed in Figs. \ref%
{fig11}(a) and \ref{fig11}(b) for $\varepsilon _{2}=+1$ and $\varepsilon
_{2}=-1$, i.e., the SF and SDF signs of the localized nonlinearity,
respectively. In the former case, the $P(k)$ dependences obey the VK
criterion, $dP/dk>0$, at $\gamma <0.28$. Accordingly, the pinned modes are
stable in direct simulations in this region, and they are destroyed by an
instability at $\gamma >0.28$, when $dP/dk$ becomes negative, see Fig. \ref%
{fig11}(a).

For the SDF sign of the localized nonlinearity, $\varepsilon _{2}=-1$, the $%
P(k) $ curves displayed in Fig. \ref{fig11}(b) satisfy the anti-VK
criterion, $dP/dk<0$. In agreement with this condition, the solitons are
found to be stable in the direct simulations.

\section*{Conclusion}

The objective of this work is to introduce the solvable model of the
nonlinear $\mathcal{PT}$-symmetric medium, in which the gain-loss
combination is represented by the point-like dipole, $\sim \delta ^{\prime
}(x)$, which is embedded into the uniform Kerr-nonlinear SF (self-focusing)
or SDF (self-defocusing) medium, in the combination with the linear and/or
nonlinear potential pinning the wave field to the $\mathcal{PT}$ dipole. The
host medium may be linear too. The full set of analytical solutions for
pinned modes has been found for this model, along with the solutions for the
system of separated $\mathcal{PT}$-symmetric point-like gain and loss sites
with the localized Kerr nonlinearity, embedded into the linear medium (the
solution for the latter variety of the solvable system makes it possible to
explicitly demonstrate the nonexistence of the $\mathcal{PT}$-symmetric
modes above the critical value of the gain-loss coefficient). The analytical
solutions were compared with numerical ones, obtained in the model with the
ideal $\delta ^{\prime }$- and $\delta $- functions replaced by their
Lorentzian regularization. It has been concluded that, with the increase of
the gain-loss-dipole strength, $\gamma $, the shape of the pinned mode
supported by the SF bulk nonlinearity gradually deviates from the analytical
limit, changing from the single-peak form into the double-peak one, which
coincides with the destabilization of the pinned soliton against escape. On
the contrary, all the pinned modes found in the model with the SDF sign of
the bulk nonlinearity are stable, featuring the single-peak shape.

The models introduced in this work can be extended in other directions. In
particular, the possibility of defining the nonlinear $\mathcal{PT}$%
-symmetry \cite{AKKZ} suggests making the gain-loss dipole nonlinear too, so
that Eq. (\ref{eq}) is replaced by 
\begin{equation}
iu_{z}=-\frac{1}{2}u_{xx}-\left( \varepsilon _{0}+\varepsilon
_{2}|u|^{2}\right) u\delta \left( x\right) +i\left( \gamma _{0}-\gamma
_{2}|u|^{2}\right) u\delta ^{\prime }\left( x\right) -\sigma |u|^{2}u.
\label{nonlin}
\end{equation}%
The corresponding stationary equation [cf. Eq. (\ref{U})] is, in principle,
solvable, although the respective algebra turns out to be cumbersome. On the
other hand, it may be interesting to introduce a two-dimensional version of
the system, with a \textit{gain-loss quadrupole} emulating the corresponding
singular expression, $\delta ^{\prime }(x)\delta ^{\prime }(y)$.

\section{Acknowledgment}

The work of T.M. was supported by the Thailand Research Fund through grant
No. RMU5380005.

\end{document}